# The role of contact resistance in graphene field-effect devices

Filippo Giubileo[a,*] and Antonio Di Bartolomeo[a,b]

[a]*CNR-SPIN Salerno, via Giovanni Paolo II 132, Fisciano (SA), Italy*
[b]*Physics Department, University of Salerno, via Giovanni Paolo II 132, Fisciano (SA), Italy*

*Corresponding author: filippo.giubileo@spin.cnr.it

## Abstract

The extremely high carrier mobility and the unique band structure, make graphene very useful for field-effect transistor applications. According to several works, the primary limitation to graphene based transistor performance is not related to the material quality, but to extrinsic factors that affect the electronic transport properties. One of the most important parasitic element is the contact resistance appearing between graphene and the metal electrodes functioning as the source and the drain. Ohmic contacts to graphene, with low contact resistances, are necessary for injection and extraction of majority charge carriers to prevent transistor parameter fluctuations caused by variations of the contact resistance. The International Technology Roadmap for Semiconductors, toward integration and down-scaling of graphene electronic devices, identifies as a challenge the development of a CMOS compatible process that enables reproducible formation of low contact resistance. However, the contact resistance is still not well understood despite it is a crucial barrier towards further improvements. In this paper, we review the experimental and theoretical activity that in the last decade has been focusing on the reduction of the contact resistance in graphene transistors. We will summarize the specific properties of graphene-metal contacts with particular attention to the nature of metals, impact of fabrication process, Fermi level pinning, interface modifications induced through surface processes, charge transport mechanism, and edge contact formation.

# Contents

# 1. Introduction

Graphene is a two-dimensional material composed of an atomically thick layer of $sp^2$-bonded carbon atoms arranged in a honeycomb structure, in which three covalent bonds are formed in the plane by the hybridization of the $2s$ orbital with the $2p_x$ and $2p_y$ orbitals with the characteristic angle of $120°$. The orbital $p_z$ is perpendicular to the plane and forms a weak π-bond. The weakness of the Van-der-Waals force between layers in the three-dimensional graphite causes the easy peel-off of single graphene sheets from the bulk material. Indeed, the single layer graphene was obtained in 2004 by mechanical exfoliation of highly oriented pyrolytic graphite, repeatedly peeling flakes by scotch tape [1]. The π-bonds are not localized and are responsible for electronic conduction properties in the graphitic structures. Graphene has been known for many years as the building block of graphite, and its electronic band structure was first calculated already in 1947 [2] within the nearest-neighbor tight-binding theory. The conduction and the valence bands are not separated by a gap, and they meet in two inequivalent points, called Dirac points, of the Brillouin zone. The Fermi level for undoped graphene lies exactly at the Dirac points, and graphene can be considered a gapless semiconductor or a zero-overlap semimetal. The electron dispersion $E(k)$ around the Dirac points is linear rather than parabolic, as in most semiconductors. The charge carriers behave as relativistic massless particles moving with an effective speed $v \approx 10^6$ m/s. Consequently, the electron transport properties are described by the Dirac's equation, with several (relativistic) quantum mechanical effects [3-7] such as the unusual half-integer quantum Hall effect [4], Klein tunneling effect [7], minimum conductivity [5], and Veselago lensing [8].

Along with its unique electronic properties, graphene has shown several interesting properties:

i) The optical absorption of single layer graphene is $A \approx 1 - T \approx \pi\alpha \approx 2.3\%$ (i.e., $T \approx 0.977$, where $\alpha = e^2/(4\pi\varepsilon_0 \hbar c) \approx 1/137$ is the fine-structure constant) [9], to be compared with the maximum visible transmittance of $T \approx 0.81$ for indium tin oxide that represents the state of the art transparent conductor. Moreover, graphene only reflects <0.1% of the incident light in the visible region;

ii) The mechanical breaking strength of defect-free single layer graphene, probed by nanoindentation in atomic force microscopy, is 42 N/m corresponding to an extraordinary Young's module of ~1 TPa [10] and intrinsic strength of ~130 GPa, confirming the graphene as the strongest material;

iii) The intrinsic thermal conductivity $K$ has been experimentally obtained for suspended graphene by optothermal Raman technique reporting $K \sim 5 \cdot 10^3$ Wm$^{-1}$K$^{-1}$ at room temperature [11]. For few layer graphene $K$ values decrease with the number of layers [12] approaching the graphite limit of ~2000 Wm$^{-1}$K$^{-1}$. The thermal properties also include the unique characteristic of a negative thermal-expansion coefficient $\alpha = -4.8 \cdot 10^{-6}$ K$^{-1}$ with a sign change at $T \approx 900$ K for single layer graphene and $T \approx 400$ K for bilayer graphene [13], as well as a very high melting point that has been estimated by atomistic Monte Carlo simulations as $T_m \simeq 4510$ K [14].

Graphene extraordinary properties have been already exploited for several applications such as gas sensors [15], photodetectors [16, 17], solar cells [18], heterojunctions [19], field-effect transistors [20,21], transparent conductors for touch screens [22,23], electromagnetic interference shielding [24], interconnects [25], flexible electronics [26], nanoelectromechanical systems [27], and antennas [28]. The electrical properties of graphene make it one of the most promising candidate for next-generation high-speed field-effect transistors. However, the contact resistance of metal/graphene interface represents a crucial limiting factor for the device's performance, affecting for instance the transconductance, the *ON/OFF* current ratio and the *cut-off* frequency.

Several studies concerning the contact resistance have reported large sample-to-sample variations. Indeed, contact resistance depends on several factors such as type of metal, substrate, self-doping, fabrication process, contact geometry. Moreover, discrepancies can be likely due also to measurement methods and/or conditions.

In this review, we first summarize the graphene properties, such as electronic band structure, carrier mobility and band gap engineering, particularly relevant for the development of high performance field effect transistors and more generally electronic devices. After that, we discuss the physics of the metal/graphene interface, the arising contact resistance and the most used measurement methods. Finally, we review the experimental and theoretical activity for improving the contact resistance via surface treatments, work function engineering, contact design, etc.

## 2. Graphene properties for field effect transistor applications

The Moore's law has correctly predicted for decades that about 18-24 months are necessary to double the number of transistors packed into an integrated chip and it has represented a continuous reference for the semiconductor industry towards successive developments of technological targets. In these years, the approach of scaling the silicon dioxide dielectric has been considered the most effective since the value below 2 nm has been reached in the 45 nm technology [29]. Few atomic layers are almost at the physical limit of the scaling process and quantum mechanical phenomena become relevant to the device performance, causing increase of the gate leakage current due to electron tunneling already at low bias and field emission at higher bias [30]. Further scaling of the $SiO_2$ produces intolerable leakage and power consumption and low breakdown voltage. A possible path towards a further scaling is the implementation of oxide materials with high dielectric constant $k$ (such as hafnium oxide, $HfO_2$ and titanium oxide, $TiO_2$) to improve the oxide capacitance, to lower the power consumption and to increase the breakdown voltage. The oxide fills the capacitor volume formed in the transistor between the gate electrode and the conduction channel, with an oxide capacitance $C_{ox} = \varepsilon_0 \cdot k / T_{ox}$ with $\varepsilon_0$ the vacuum permittivity and $T_{ox}$ the oxide thickness. For high-$k$ materials an equivalent oxide thickness $T_{eq}$ corresponding to the $SiO_2$ thickness necessary to obtain the same capacitance can be easily estimated as $T_{eq} = \left( \frac{\varepsilon_{SiO_2}}{\varepsilon_{high-k}} \right) \cdot T_{high-k}$.

Different routes to improve the device performance of the silicon technology point towards three dimensional devices and/or uniaxial strained silicon channels. A great effort is focused on the search for new devices to replace the silicon based field-effect transistor (FET) [31] by using new channel materials. Indeed, channel thickness ($T_{channel}$) and gate oxide thickness are fundamental quantities to eliminate short-channel effects because the characteristic channel length is proportional to $\sqrt{T_{ox} \cdot T_{channel}}$ [32]. This makes thinner channel materials suitable for transistor design.

Actually, several attempts have been already reported to realize innovative FETs, to keep the requirements of modern electronics technology, using III–V compound semiconductors (such as GaAs, AlAs, InAs, InP) [33], silicon nanowires [34], carbon nanotubes [35], organic materials [36], graphene [20] and other 2-dimensional materials such as transition metal dichalcogenides ($MoS_2$, $WS_2$,…) [37], and monoatomic buckled crystals including silicene [38], germanene [39] and phosphorene [40].

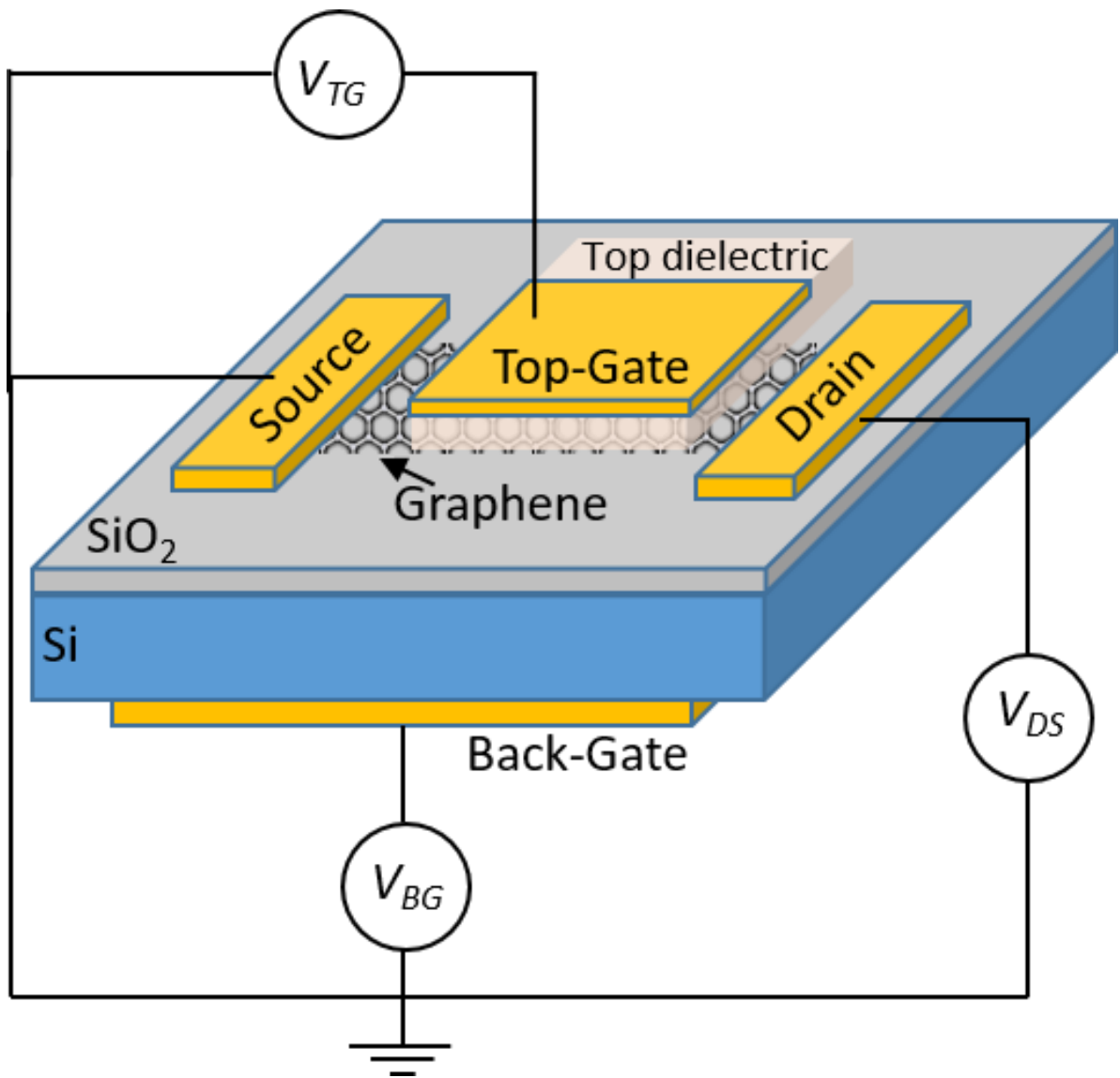

**Fig. 1.** Schematic of a field effect transistor with graphene as transistor channel between source and drain electrodes. In a standard configuration the graphene flake is placed on a heavily doped Si/SiO₂ substrate. The silicon substrate acts as back-gate being separated by 300 nm thick $SiO_2$ from the channel. A second dielectric layer on top of graphene allows to have a top-gate electrode. $V_{DS}$ is the applied bias between source and drain, while $V_{BG}$ and $V_{TG}$ are the back-gate and top-gate bias, respectively.

A field effect transistor is a three-terminal device in which the current flows along a (*n*-type or *p*-type) semiconductor channel contacted between two metal electrodes, the source (S) and the drain (D). The third electrode is named gate (G) and it is located in close proximity of the channel, forming a capacitor (see Fig. 1). A FET allows to control the channel conductivity by means of a gate voltage that capacitively induces an electric field on the channel, modifying the carrier density and consequently the conductivity. There exist several types of FET depending on how the gate capacitor is realized (JFET, FINFET, MOSFET). In a metal-oxide-semiconductor FET (MOSFET) the gate is insulated from the channel by a thin oxide layer (normally silicon dioxide or similar). The conventional silicon based MOSFET has several characteristics such as an high *ON/OFF* ratio. It can operate in the i) *cut-off* region, in which the gate voltage is less than the threshold voltage required for conduction, so that there is negligible current in the channel, $I_{OFF}$; ii) linear region, in which the channel conduction is linearly controlled by the gate voltage; iii) saturation region, in which the drain current $I_{ON}$ is only weakly dependent on drain voltage while is mostly controlled by gate-source voltage.

The unique band structure and the high carrier mobility make graphene an extraordinary alternative to silicon for FET applications, also due to the planar geometry, suitable for processing within the standard complementary metal oxide semiconductor technology. One of the principal difficulties for the exploitation of graphene in this field is certainly related to the absence of bandgap that prevents a graphene based FET (GFET) to be completely switched off. A possible solution is to open a bandgap by introducing the use of nanoribbons or bilayers.

The back-gated configuration represents the most analyzed GFET structure, in which a graphene flake is placed on a substrate (typically $Si/SiO_2$ with oxide 90 nm or 300 nm thick) and is contacted between source and drain electrodes, while the substrate (Si) is the back gate. It is possible to realize a further insulating layer on top of the device, in order to place also a top-gate electrode, both gates controlling the carrier concentration in the channel.

Due to the zero bandgap of graphene, the GFET channel has a low current ratio $I_{ON}/I_{OFF}$, that at room temperature doesn't reach 10. This strongly limits the possible exploitation of graphene in logic device applications, a minimum ratio $10^3$ being requested to properly approximate the complete switch off of the transistor in the *OFF*-state (almost no current flowing in the channel). On the other hand, enormous interest is addressed to the use of graphene in the field of analog circuits, because they are normally biased in the *ON*-status making the ratio $I_{ON}/I_{OFF}$ less relevant, with respect the current gain [41]. Graphene-based radio frequency field-effect transistors with *cut-off* frequency (which is the maximum frequency at which a gain is achievable) above 400 GHz have been already demonstrated [42].

## 2.1. Band structure

The honeycomb structure of graphene is realized by the three strong covalent $\sigma$-bonds that each atom forms in plane at a distance of $a = 1.42$ Å, while one electron per atom forms the $\pi$-bonds, the electronic properties of graphene at low energies being due to the $\pi$-electrons. On the contrary, the $\sigma$ electrons are responsible for energy bands far from the Fermi energy. The graphene crystal structure (see Fig. 2) may be seen as a triangular Bravais lattice with a two-atom basis (an atom for each sublattice, the so-called A and B sublattices): $\boldsymbol{a_1} = a\left(\sqrt{3}/2, 1/2\right)$ and $\boldsymbol{a_2} = a\left(\sqrt{3}/2, -1/2\right)$. The reciprocal lattice vectors are $\boldsymbol{b_1} = 2\pi/3a\left(1, \sqrt{3}\right)$ and $\boldsymbol{b_2} = 2\pi/3a\left(1, -\sqrt{3}\right)$.

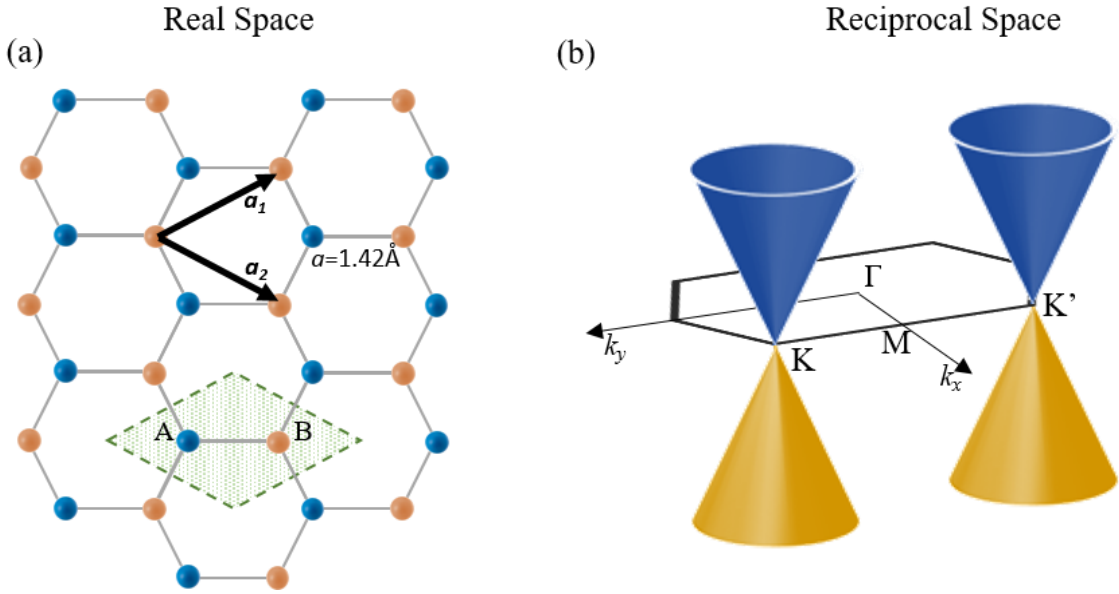

**Fig. 2.** (a) The honeycomb lattice structure of graphene with two atoms, A and B, per unit cell. $\boldsymbol{a_1}$ and $\boldsymbol{a_2}$ are the lattice vectors and $a = 1.42$ Å is the in-plane nearest-neighbour distance. (b) Brillouin zone of graphene with representation of Dirac cones near the K and K' points.

The first Brillouin zone is hexagonal, with two inequivalent points $K$ and $K'$ at the six corners, the position being expressed as $K = 2\pi/3a\left(1, 1/\sqrt{3}\right)$, $K' = 2\pi/3a\left(1, -1/\sqrt{3}\right)$. The electronic band structure of graphene has been calculated within the nearest neighbor tight binding [2] approximation, considering only hopping of electrons between nearest-neighbor atoms from sub-lattice A to B or viceversa. This is a reasonable approximation for small wavevectors around the $K$ and $K'$ points [43]. The dispersion relation $E(\vec{k})$ can be obtained from the Hamiltonian

$$H = -t \sum_{\langle i,j \rangle, \sigma} \left(a_{\sigma,i}^+ b_{\sigma,j} + H.c.\right)$$

where $a^+_{\sigma,i}$ creates an electron on the sublattice A with spin $\sigma$ (up or down) on the $i$-th site, while $b_{\sigma,j}$ annihilates an electron on the sublattice B, and $t \approx 2.8$ eV is the hopping energy. The resulting energy bands are expressed as

$$E(\vec{k}) = \pm t\sqrt{1 + 4cos(\sqrt{3}ak_x/2)cos(ak_y/2) + 4cos^2(ak_y/2)}$$

The positive (negative) sign refer to the upper (lower) band, usually referred as $\pi$-band ($\pi$*-band). For undoped graphene, the conduction and the valence band meet at the $K$ and $K'$ points (*Dirac points*) of the Brillouin zone, and this makes graphene a semiconductor with zero bandgap. Close to the Dirac points the dispersion relation is linear and it can be written as

$$E = \pm \hbar v_F |\boldsymbol{\kappa}|$$

with $\boldsymbol{\kappa} = \vec{k} - \vec{K}$ the wavevector measured relatively to the Dirac point and $v_F \approx 10^6$ m/s the Fermi velocity. The linear dispersion, in contrast to the usual parabolic dispersion relation with a mass dependence ($E = \hbar^2 k^2/2m$), indicates that the charge carriers move with a velocity that is independent of the energy, thus behaving as relativistic and massless particles (Dirac fermions) with velocity $v_F$. This is one of the most intriguing properties, making graphene the object of huge scientific effort worldwide. The Fermi energy can be written in terms of carrier density $n$ as $E_F = \hbar v_F \sqrt{\pi n}$ which implies that it is possible to tune the Fermi energy by changing the carrier density (either electrons or holes) via a bias gate in a graphene based field effect transistor. The density of states $D(E)$ in graphene is easily obtained: $D(E) = 2E/(\pi \hbar^2 v_F{}^2)$. Thus, $D(E)$ vanishes linearly at the Dirac point, differently from the two dimensional electron gases with parabolic energy dispersion relation for which the density of states is energy independent.

In condition of thermal equilibrium, pristine graphene has both mobile electrons and holes, with identical intrinsic carrier concentration

$$n_i = n = p \simeq \frac{\pi}{6}\frac{(kT)^2}{\hbar^2 v_F{}^2}$$

where $n_i \cong 8.2 \cdot 10^{10}$ cm$^{-2}$ for $v_F = 10^6$ m/s at room temperature. The electron (hole) concentration $n$ ($p$) in a graphene layer is given by

$$n = \int_0^\infty \frac{2|E|}{\pi\hbar^2 v_F{}^2} f(E)dE$$

and

$$p = \int_{-\infty}^0 \frac{2|E|}{\pi\hbar^2 v_F{}^2} \big(1 - f(E)\big)dE$$

with $f(E) = \big(1 + e^{(E-E_F)/kT}\big)^{-1}$.

Electrons or holes can be majority carriers in graphene, depending on the position of the Fermi energy with respect to the Dirac point (see Fig. 3). The control of the Fermi level can enable the switching from *n*- to *p*-dominated transport (ambipolar effect). From these expressions, the carrier density is zero at T=0K. For T>0K, there is always a non-zero carrier density.

Morphological corrugations (ripples) and defects make graphene sheets not perfectly planar (thus favoring its thermodynamic stability [44]); potential fluctuations associated to them or to charges in the surrounding environment induce the formation of electron (hole) puddles, i.e. regions with enriched number of electrons (holes) in neutral graphene [45].

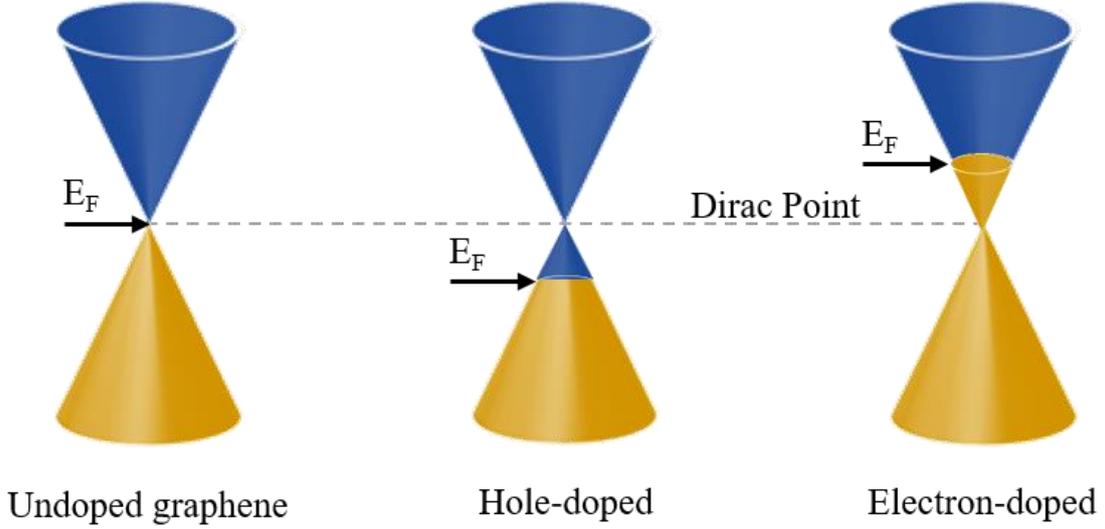

**Fig. 3.** Representation of the linear dispersion (Dirac cone) with conduction and valence bands touching at the K point (Dirac point). Charge transfer to (from) graphene causes electron (hole) doping with the Fermi level displaced above (below) the Dirac point.

Near the Dirac points the low energy physics of graphene is well understood in terms of two component wavefunctions corresponding to the particle density of the two sublattices (A and B):

$$\psi_{\pm,K}(\vec{k}) = \frac{1}{\sqrt{2}}\begin{pmatrix} e^{-i\theta_{\vec{k}}/2} \\ \pm e^{i\theta_{\vec{k}}/2} \end{pmatrix}$$

with $\theta_{\vec{k}} = arctan\left(\frac{k_x}{k_y}\right)$ and the signs corresponding to the $\pi$* (+) and the $\pi$ (-) bands. Such wavefunction has the important consequence that for a complete rotation of $2\pi$ of the momentum at constant energy around a Dirac point, the wavefunction undergoes a phase change of $\pi$ (Berry's phase) [7] instead of an expected 0 or $2\pi$. This property is strictly related to magneto-transport phenomena (such as unconventional quantum Hall effect), being the mentioned rotation usually produced by applying an external magnetic field.

The two-component wavefunction used to describe the graphene has a pseudospin that refer to the sublattice (A or B) rather than to a real spin of the electrons, and the conservation of pseudospin implies that backscattering in graphene is prohibited [46], originating fascinating phenomena such as the Klein tunneling.

The pseudospin allows the introduction of a chirality [47], defined as the projection of pseudospin on the direction of motion. Indeed, electrons and holes belonging to the same branch of the electronic spectrum have pseudospin in the same direction, parallel to the momentum for electrons and antiparallel for holes.

## 2.2. Transport properties: mobility and saturation velocity

Depending on the mean free path $\ell$, two possible transport regimes can be realized in graphene devices of length $L$. Real graphene is normally placed on a substrate, which may induce defects and sheet corrugations (ripples). All these conditions modify the electronic properties of ideal graphene causing either carrier density inhomogeneities (puddles) and/or reduction of $\ell$. For $L < \ell$ transport in graphene is ballistic, with the charge carriers travelling with velocity $v_F$ and conductivity expressed in the Landauer formalism in terms of the transmission probabilities $T_n$ of each possible transport mode [48]:

$$\sigma_{Bal} = \frac{4e^2}{h} \frac{L}{W} \sum_{n=1}^{\infty} T_n$$

The minimum conductivity at the Dirac point is theoretically calculated as:

$$\sigma_{min} = \frac{4e^2}{\pi h} \simeq 4.9 \cdot 10^{-5} S$$

On the contrary, the diffusive regime in graphene is realized when $L > \ell$, with charge carriers experiencing scattering in their motion, and it can be described by means of semiclassical transport theory (Boltzmann). The conductivity in this regime is expressed at low temperature as

$$\sigma_{Dif} = \frac{2e^2 v_F}{h} \tau \sqrt{\pi n}$$

with $\tau$ the total relaxation time that is related to the scattering mechanisms, such as Coulomb scattering due to charged impurities, short range scattering due to vacancies in the graphene layer, or phonon scattering.

One of the most important electronic properties of graphene is surely the charge carrier mobility (μ), being a fundamental performance parameter that influences current intensity and frequency response in transistors. By definition, $\mu \equiv v/E$, with $v$ the carrier drift velocity and $E$ the electric field, and it can be written as $\mu = \sigma/en$, where $\sigma$ is the conductivity. The mobility can be estimated through magnetoresistance measurements in Hall-bar configuration ($\mu_H$) as well as via resistivity measurements versus gate bias in graphene based field effect transistors ($\mu_{FE}$). This last approach allows two-probe measurements, avoids any special setup to apply magnetic field and gives almost the same result of Hall-bar experiment, i.e. $\mu_{FE} \approx \mu_H$.

According to theoretical expectations, graphene may have high intrinsic carrier mobility (either for electrons and holes) [49], the only limiting factor being the phonon scattering [50]. Experimentally, highest mobility values have been reported for suspended graphene at low temperature (~5K) as ~$2 \cdot 10^5$cm$^2$V$^{-1}$s$^{-1}$ [33].

By increasing temperature, the mobility is linearly reduced due to the increased scattering that depends on the acoustic phonons. However, approaching room temperature, the mobility can still be very high ($\mu > 10^5$ cm$^2$V$^{-1}$s$^{-1}$, corresponding to a mean free path $\ell \approx 1\mu$m [50]) with respect the electron mobility in Silicon ($\mu_{Si} \simeq 1.5 \cdot 10^3$ cm$^2$V$^{-1}$s$^{-1}$) or III-V compound semiconductors such as GaAs or InAs ($\mu_{GaAs} \simeq 8.5 \cdot 10^3$ cm$^2$V$^{-1}$s$^{-1}$, $\mu_{InAs} \simeq 33 \cdot 10^3$ cm$^2$V$^{-1}$s$^{-1}$). The III-V compound semiconductors have also very limited hole mobility [33].

Despite the impressive mobility reported in graphene, the use of suspended layers poses severe limitations to the design of graphene based devices for large scale applications. On the contrary, the use of substrates to sustain graphene has detrimental effect on the mobility. For example, for exfoliated graphene placed on SiO$_2$ substrate, low field carrier mobility below $1.5 \cdot 10^4$cm$^2$V$^{-1}$s$^{-1}$ is always reported [51, 52], while for epitaxial graphene (or chemical vapor deposited graphene) on SiC substrate the upper limit is $10^4$cm$^2$V$^{-1}$s$^{-1}$. The low density of defects and the large area of the exfoliated graphene flakes make such form the one with highest quality (for instance respect to CVD-graphene, i.e. grown by Chemical Vapor Deposition), allowing the high carrier mobility, a good indicator of high quality.

Another possible substrate for graphene is the hexagonal boron-nitride (h-BN) that has been proposed [53] to exploit several advantages (such as a lattice constant, similar to graphene, an atomically flat surface with very few charged impurities, high surface phonon frequencies) in order to obtain a carrier mobility three times higher than using SiO$_2$ substrates. Moreover, it has been reported that for graphene encapsulated in h-BN the transport is ballistic at room temperature for distances greater than 1$\mu$m and the mobility is greater than $10^5$ cm$^2$V$^{-1}$s$^{-1}$ [54].

Despite the very high carrier mobility, if considering short channel GFETs, high electrical fields are achieved (~$10^5$V/cm), and the saturation velocity become a relevant parameter, that in graphene ($v_{sat} \approx 4 \cdot 10^7$cm/s) is higher than in other semiconductors ($1 \cdot 10^7$cm/s for Si and $2 \cdot 10^7$cm/s for GaAs). Experiments suggest that, in the high-bias regime, optical phonons of graphene (~160 meV) [55,56] and SiO$_2$ (55 meV) [57,58] are both relevant to limit the saturation velocity [59]. At room temperature, $v_{sat}$ decreases for rising temperature and for increasing carrier density above $2 \cdot 10^{12}$ cm$^{-2}$, and appear limited by the substrate [59]. Current saturation has been measured in dual-gated short-channel GFETs on Si/SiO$_2$ and it has been modeled within a velocity saturation model of high-field transport [60] that predicts a dependence from the carrier density as $v_{sat} \sim n^{-1/2}$.

Transport experiments in the high-current regime of graphene based devices suggest the possibility of an incomplete saturation, because of the competition between elastic-scattering (disorder) and optical phonon scattering [56] or due to the formation of a pinch-off region [60]. The complete understanding of the mechanisms limiting the saturation velocity is still lacking.

### 2.3. Zero bandgap and Bandgap opening in graphene

One of the major limitations to the use of graphene in digital electronics is the absence of an energy bandgap. A semiconductor with zero energy gap as graphene can be properly exploited for high-speed analog electronics and transparent conductive films [22], but is not suitable for logic applications, where a complete switch-off of GFET is requested in order to have acceptable ($> 10^3$) current ratio $I_{ON}/I_{OFF}$ with respect a typical ratio $\sim 5 \div 10$ in GFETs on Si/SiO$_2$.

Because the zero bandgap in graphene is due to the presence of two identical carbon atoms in the unit cell [61], in order to open a sizable bandgap in graphene (i.e., to realize semiconducting graphene) it is necessary to break the planar symmetry of the crystal structure by means of structural and/or chemical modifications. For instance, the substitution of a carbon atom with nitrogen will break the in-plane symmetry of the hexagonal lattice inducing the formation of energy gap between the $\pi$ and $\pi$* bands [62]. Graphene-substrate interactions have been also proposed as methods to open an energy bandgap in graphene [63,64]. For instance, the use of SiC substrate can cause a bandgap opening ($\sim 0.26$ eV) because the lattice mismatch with graphene that breaks the sublattice symmetry [63].

A gap opening can be also obtained by breaking the symmetry along the vertical (c-) axis stacking two layers, the AB-stacking, to form a bilayer graphene (BLG) [65]. This is because the unit cell of BLG has four atoms in the unit cell, originating two additional bands and in presence of strong electric field perpendicular to layers a gap between the lower conduction band and the higher valence band is opened. In one of the first studies, the carrier concentration in each layer of a BLG on SiC substrate was separately modified in order to open and control an energy band gap up to a maximum value of 0.2 eV [66]. The possibility to tune the energy bandgap in BLG based FETs via opportune electrical gating has been demonstrated [67] reporting continuously tunable bandgap up to 0.25 eV. In dual-gated GFET on Si/SiO$_2$, current ratio $I_{ON}/I_{OFF} \sim 10^2$ and energy gap $\sim 0.13$ $eV$ has been obtained [68] at room temperature for an average electrical displacement of 2.2 V/nm, and $I_{ON}/I_{OFF} \sim 2 \cdot 10^3$ at low temperature (T = 20 K). The energy gap in BLG can be created also by molecular doping: organic triazine has been used to produce a thin film covering the BLG, protecting the top layer from ambient *p*-doping [69]. This results in a bandgap opening up to 0.11 eV and $I_{ON}/I_{OFF} \sim 60$ at room temperature. It has also been demonstrated that the gap opening in BLG can be obtained also in single gate configuration by dual doping, i.e. realizing *n*-doping from the bottom side and *p*-doping from the top side [70].

Another possibility of opening a gap in graphene is represented by the quantum confinement that is realized in one-dimensional graphene nanoribbons (GNR), in graphene nanomesh or graphene quantum dots. Indeed, as consequence of the patterning, graphene lattice cannot be approximated as semi-infinite plane and the charge carries result laterally confined [71] producing at the Fermi level a bandgap inversely proportional to the GNR width. GNRs may be produced by means of lithographic techniques, a method favorable to obtain desired geometries but it has intrinsic limitation with respect the size scaling due to the technique resolution ($\sim$20 nm) and also it produces GNRs with limited edge quality that affects the device performances [71]. Actually, several fabrication methods, other than e-beam lithography, have been developed to produce GNRs, such as unzipping of CNTs by plasma etching [72,73], nanowire lithography [74] or by applying a gas phase etching step after the lithography [75]. In all cases GNRs less than 10 nm wide are obtained, with energy band gap up to 0.5 eV [73] and current ratio $I_{ON}/I_{OFF}$ up to $10^4$ [75].

In the graphene nanomesh structures, a dense array of holes are realized on a graphene sheet, the electronic properties strongly depending on array parameters (width of holes and separation distance) [76]. Current ratio $I_{ON}/I_{OFF}$ up to $10^2$ [77] and bandgap below 0.2 eV [78, 79] are reported.

For graphene quantum dots, produced by electron beam lithography, with lateral dimensions as small as $\sim$15 nm, it has been reported the opening of energy gap up to $\sim$0.5 eV [80].

Another simple method to open a gap is by realizing a complete oxidation of graphene, the so called graphene oxide (GO), practically an insulator with a band gap of 2.1 eV, due to the $sp^3$ hybridization of carbon atoms bonded with the oxygen [81]. GO has interesting electronic properties that can be tuned starting from a sheet resistance as high as $10^{12} \Omega/\square$ for the insulating state [82]. A reduction process, by chemical or thermal treatments, can modify the ratio between the $sp^2$ and $sp^3$ carbon fractions allowing the tuning of the bandgap corresponding to an evolution of the GO towards semiconducting state and even to graphene-like zero-gap state [83]. The controlling of the coverage, and/or arrangement and/or relative ratio of the epoxy and hydroxyl groups is also a suitable method to tune the GO bandgap [84-87].

## 3. The role of metal contacts

When a contact between a metal (M) and a semiconductor (S) is realized, two possibilities can originate: i) an ohmic contact, generally obtained with heavily doped S, in which current can flow in both directions and it is linearly dependent on the applied voltage (the total resistance being the sum of the contact resistance R and the bulk resistances of the materials); ii) a rectifying Schottky diode, normally obtained on lightly doped S, that allows the current flow easily only in one direction [88]. The rectifying behavior arises from a potential barrier (Schottky barrier) forming at the interface. Semiconducting devices or more generally integrated circuits need ohmic contacts with other electronic systems for proper operations.

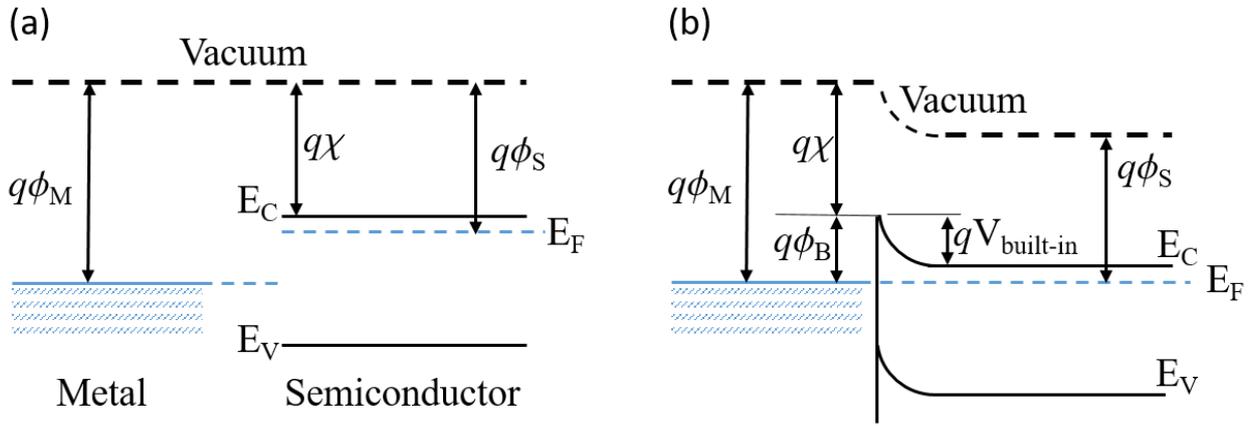

**Fig. 4.** Energy band diagram of (a) metal adjacent to an isolated *n*-type semiconductor, (b) metal-semiconductor contact in thermal equilibrium.

The energy band diagrams, reported in Fig. 4(a), show a metal with workfunction $\phi_M$ , i.e. the energy difference between the Fermi energy and the vacuum level, and a semiconductor, whose workfunction $q\phi_S = q(\chi + V_n)$ is written in terms of electron affinity $q\chi$ and $V_n = E_C - E_F$ the difference between the bottom of the conduction band and the Fermi level, initially separated. When M and S are brought into intimate contact, a single system (Fig. 4(b)) is formed and it reaches the thermal equilibrium in which the alignment of the Fermi levels on both sides is realized, due to carriers flowing from higher energy states to lower energy states. The small density of states (DOS) in S causes band bending in the so-called depletion layer. In the ideal case, the barrier height $q\phi_B$ can be expressed as $q\phi_B = q(\phi_M - \chi)$. The electrons moving from (*n*-type) S to M feel a so-called built-in potential $V_{built-in} = \phi_B - V_n$. In equilibrium, the current flow from S to M and viceversa are equal. Under forward bias, electron flow from S to M is enhanced because a lowered potential barrier, while the opposite flow from M to S is unchanged, $\phi_B$ remaining the same. Under reverse bias, the barrier is increased, and the electron flow from S to M is negligible, the reverse current being due to electrons moving from M to S over $\phi_B$. Thus, the Schottky barrier is not modified by voltage bias, unless second order effects, and this means that it can be controlled by opportune material choice.

The Schottky diode is an unipolar device because the minority carrier current (holes injection from M to n-type S) is much smaller than majority carrier current and it is obtained in M/S contacts with high barrier ($\phi_B \gg kT$). In such a case, the current transport at room temperature is mostly due to thermionic emission of majority carriers from S to M and it is written as

$$I = AA^*T^2 e^{-\phi_B/kT}\left(e^{qV/kT} - 1\right)$$

with A the junction area and A* the Richardson constant [19].

There are two ways to favor an M/S contact to be ohmic: lower the barrier height or make the barrier very narrow. If using heavily doped S, the depletion layer is very thin, of the order of tens of Angstrom, and electrons can go across the barrier due to the quantum tunneling effect. This results in a linear current-voltage characteristics with a contact resistance that depends on doping concentration $N_D$ as $R_c \sim e^{\phi_B/\sqrt{N_D}}$. For low doping concentrations, the current through M/S interface is mostly due to thermionic emission and in order to get small contact resistance a low barrier height is required.

If two metals, M-1 and M-2, are brought into contact (see Fig. 5(a) and (b)), no energy barrier will arise at the interface, although they could have different workfunctions $q\phi_{M-1}$ and $q\phi_{M-2}$, respectively. Indeed, thermal equilibrium requires aligned Fermi levels on both sides. Assuming $\phi_{M-1} > \phi_{M-2}$, electrons will flow from M-2 through the interface, causing positive (negative) space charge in M-2 (M-1) near the interface. However, the space charge imbalance is screened by the high carrier density in metals on very short length ($< 1$ nm) that means an abrupt vacuum level change localized at the interface.

### 3.1 Metal-graphene interface

Much complex is the situation when interfacing a metal with graphene (G), because of the zero energy bandgap and the vanishing DOS at Dirac point (see Fig. 5(c) and (d)). The absence of energy gap prevents the formation of depletion layer and of conventional Schottky contacts. The current injection from M to the graphene is strongly limited by the small DOS near the Dirac point. Due to its two-dimensional nature, graphene is highly sensitive to environment, and its properties are strongly influenced when creating a contact with a metal. Different work functions cause a charge transfer through the interface, originating electrical doping of graphene. The small density of states in graphene near the Dirac energy is responsible of significant shift of the Fermi level in graphene even for limited charge transfer. The shift $\Delta E_F$ of the Fermi level can be upwards when electrons moves from M to G (*n*-doping) while $E_F$ shifts downwards when positive carriers (holes) move from M to G (*p*-doping).

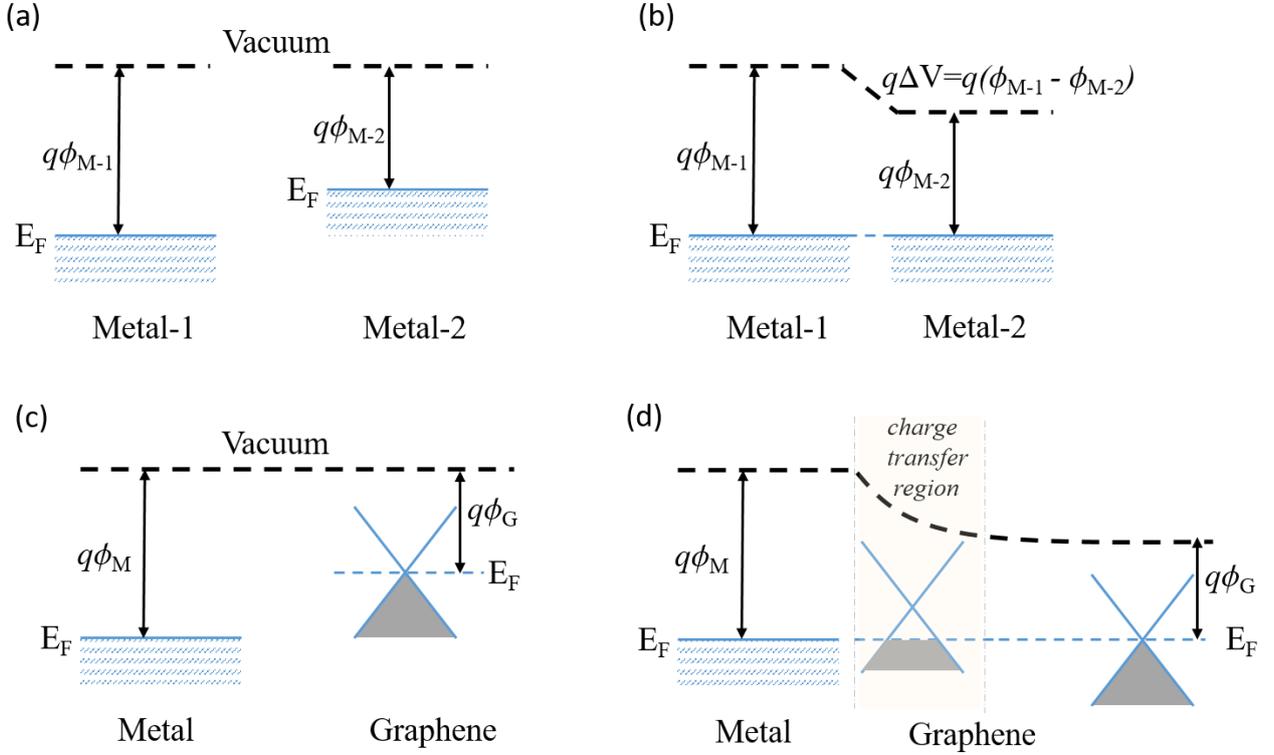

**Fig. 5.** Schematic of energy band diagram for the metal-metal and metal–graphene contacts. (a) Representation of separated metals with their own workfunctions; (b) when the two metals are brought in contact, equilibrium is reached when the Fermi levels are lined up by the transfer of electrons. A contact potential is formed at the interface. (c) Representation of separated metal and graphene (with its Dirac cone). (d) When metal and graphene are brought in contact, the Fermi levels are aligned. Far away from the M/G contact, the conical point of graphene approaches $E_F$.

In principle, a crossover from *n*-doping to *p*-doping is expected when both sides of the junction have the same doping type but different workfunction. Actually, when one of the material is graphene, the work function difference is not the only parameter, because the chemical interaction at the surface cannot be disregarded. First-principle calculations at the level of density functional theory have shown that the graphene electronic properties can be modified either by physisorption for contacting metals such as Au, Cu, Pt or by chemisorption for Ni, Co, Pd [89] at the M/G interface. By physisorption, the graphene electronic structure is only weakly altered, due to the weak binding, and the interaction causes charge transfer (depending on the workfunctions) and the shift of the Fermi level. It results that the crossover from *n*- to *p*-doping is expected for metals with workfunction ∼5.4 eV, higher than the value for free-standing graphene $q\phi_G \approx 4.5$ eV [90]. On the contrary, chemisorption causes a strong perturbation of the electronic structure of graphene due to hybridization between graphene $p_z$ states and $d$-states in the metal that opens a band gap in graphene and significantly reduces its workfunction [90]. Physisorption or chemisorption metals depend on the degree of filling in the d-orbitals, which fixes in the hybridization the stability of the antibonding states, a big number of electrons in the antibonding states destabilizing the hybridization [91].

When considering a real system in which a finite metal electrode is deposited to cover a part of a graphene sheet (as for transport measurements), the Fermi level (FL) will be at the Dirac point only in the free graphene, far away from M/G interface, where instead the FL is fixed by the metal. In order to accommodate the FL difference, there is a charge transfer between the two graphene regions (the free one and the contacted one). The area in which the band bending is realized can be $n$-doped or $p$-doped depending on the FL difference. This implies that by opportune choice of metal electrodes, it is possible to engineer $p$-$n$ junctions in graphene [90].

At the contact region, a metal induced electrostatic potential is formed, the screening in graphene being strongly suppressed with respect metals. Using a Thomas-Fermi approach to study the band bending in graphene due to the presence of metal contact [92] it has been found that there is a long-range potential that weakly decays with the distance $d$ from the interface as $\sim d^{-1/2}$ in undoped graphene, while decaying as $\sim d^{-1}$ for doped graphene, the long screening length originating from the small DOS at the Fermi level. The formation of this space charge region has been experimentally measured by scanning photocurrent microscopy [93] evidencing that the electronic structure is modified not only in the graphene under the metal, but the perturbation extends up to 500 nm in the graphene sheet far from the interface [94].

### 3.2 Metal-graphene contact resistance

The major limitation to the complete exploitation of graphene properties in electronic devices is represented by the contact resistance $R_C$ arising at the M/G interface, where complex transport phenomena take place, charge carriers being injected from three-dimensional metal electrode to two-dimensional graphene layer. In order to profit from the exceptional intrinsic properties of graphene, it is necessary to minimize the contact resistance towards the standard obtained for silicon based MOSFET technology in which contact resistivity is about 50 $\Omega\mu m$ well below the transistor resistance in the ON-state ($\sim 1/10$). In short channel GFETs, the total resistance $R_{total}$ is dominated by $R_C$ because it does not reduce with length. Also in high-speed applications, $R_C$ is detrimental on *cut-off* frequency, extrinsic transconductance, maximum frequency of oscillation, current-voltage linearity and absolute intensity of ON-state current.

Several parameters are generally used to characterize the contact resistance. With $R_C$ we refer to the contact resistance (measured in $\Omega$), while the graphene sheet resistance is indicated as $R_{sh}$ (in $\Omega/\square$), the specific contact resistivity is denoted as $\rho_C^{specific}$ (in $\Omega\mu m^2$) and the contact resistivity $\rho_C = R_C \cdot W$ (in $\Omega\mu m$) where $W$ is the contact width. In general, if the current flowing under the contact is uniform, $R_C$ can be expressed as $R_C = \rho_C^{specific} W^{-1} D^{-1}$ with $D$ the contact length. When graphene is involved in the contact, it has been demonstrated [95], by four-probe measurements on GFETs with several contacts with fixed width and different areas, that $\rho_C$ remains almost constant while $\rho_C^{specific}$ increases with the contact area. This indicates that the contact resistance arising at the graphene/metal interface depends on the contact width $W$ instead of the contact area ($W \cdot D$), confirming that at the edge of the metal contact there is current crowding [96]. Such effect is present in the limit that the contact dimension (length) $D$ is larger than the transfer length $L_T$, where $L_T$ is defined as the effective contact length contributing to the injection of carriers in graphene. If $D < L_T$, a transition from edge-to area-conduction regime will happen, with the whole contact area participating in the carrier injection.

For the edge-conduction regime (in the case $D > L_T$), the contact resistance is expressed in terms of the effective contact area $W \cdot L_T$ as $R_C = \rho_C^{specific} W^{-1} L_T^{-1}$ where $L_T = \sqrt{\rho_C^{specific}/R_{sh}}$.

For a correct evaluation of $L_T$, one should take into account that $R_{sh}$ is different in the channel and underneath the metal contact. Indeed, it has been reported that $R_{sh}$ under the contact is strongly dependent on the deposition process [97]: very high contact resistivity has been measured ($\rho_C \approx 10^9$ $\Omega\mu m$) for graphene sheet contacted by rf-sputtered Ti electrodes, while values from three to six order of magnitude less are obtained for thermally evaporated Ti/Au contacts. The high value obtained for sputtered Ti is ascribed to a large number of defects produced by the fabrication process at the M/G interface, as confirmed by the presence of D-band in the Raman spectra with respect the defect-free spectra where D-band is not observed [97].

In order to characterize the intrinsic performance of a GFET, it is necessary to evaluate the series parasitic resistances that can be five to ten times larger than the channel resistance. Several methods can be applied to measure the contact resistance: transfer length method (TLM), four-probe, cross bridge Kelvin (CBK). The most used for graphene based device is the TLM method, in which several (identical) contacts are differently spaced, as represented in Fig. 6(a). This method allows a complete characterization of contact resistance, sheet resistance and specific contact resistivity.

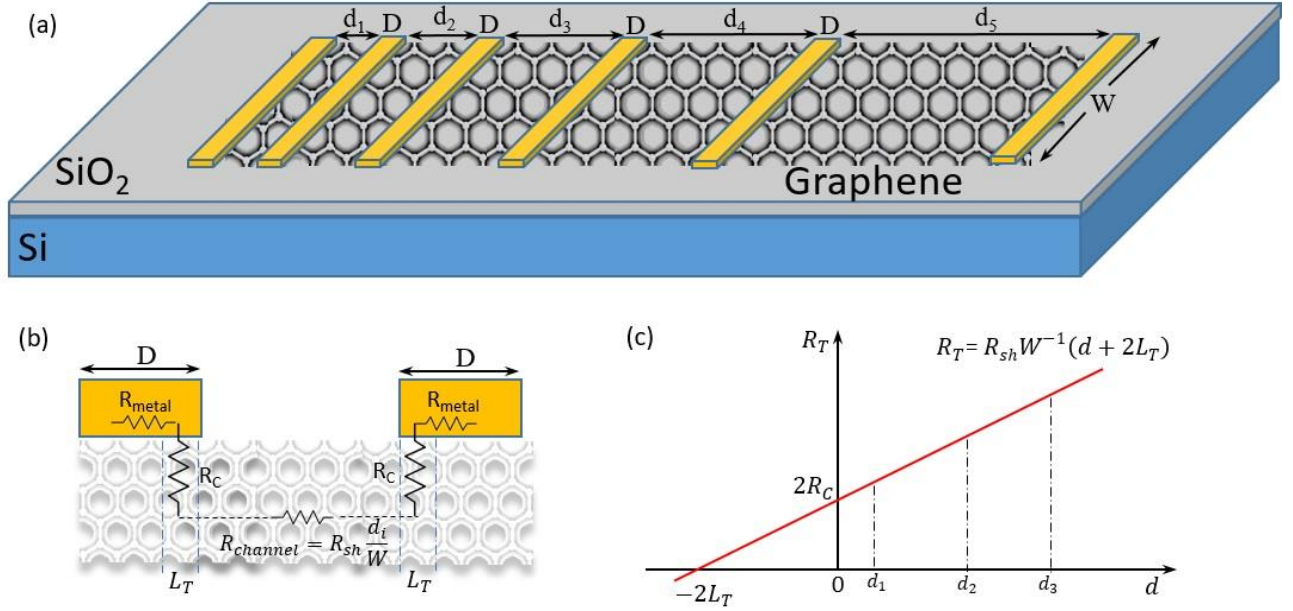

**Fig. 6.** (a) Schematic representation of a back-gated device with TLM configuration. (b) Schematic of two-contact measurements: The metal electrode resistance $R_{metal}$ is much lower that $R_C$ (mostly limited in a $L_T$ wide layer for the crowding effect) and $R_{channel}$. (c) Measurements are performed across contacts with increasing separation distance to create a curve of resistance versus distance: The intercept gives twice the contact resistance, while the sheet resistance can be derived from the slope.

For any couple of contacts, the total resistance $R_T$ is given by $R_T = R_{channel} + 2R_C$ where we are neglecting the metal resistance $R_{metal}$, the channel resistance between two successive contacts being $R_{channel} = R_{sh} \frac{d_i}{W}$ with $d_i$ the separation between the considered contacts. In the approximation that the contact width $W$ is equal to the sample width, and the contact length is greater than $L_T$, the contact resistance is expressed as $R_C = \rho_C^{specific} W^{-1} L_T^{-1}$ and by using $L_T = \sqrt{\rho_C^{specific}/R_{sh}}$ the total resistance can be written as $R_T = R_{sh} W^{-1}(d_i + 2L_T)$. Consequently, by measuring $R_T$ as a function of contact separation $d_i$ will result in a linear plot, whose intercept at $d_i = 0$ corresponds to the value $R_T = 2R_C$, while the intersection with the horizontal axis ($R_T = 0$) corresponds to the value $-2L_T$ giving an estimation of the transfer length (Fig. 6(b) and (c)). Finally, the slope of the linear function $R_T$ vs $d_i$ is $m_{TLM} = R_{sh} W^{-1}$ and it allows to evaluate the sheet resistance.

Although the transfer length method is widely used to measure the contact resistance in GFETs, it has been often disregarded that the standard method applied for conventional semiconductors (M/S contacts) is based on two assumptions not obviously realized when graphene is involved in the contact (M/G): first, the sheet resistance is the same underneath the contact and in the channel, second, the contact is diffusive due to a very short electron mean free path in doped S. A more reliable model needs to consider the different sheet resistance in the channel and under the contact, due to the metal doping of graphene, as introduced by Xia et al. [98]. According to the model, the charge carrier transport is realized by a two-step process: the carrier injection from the metal to the graphene underneath the contact with probability $T_{MG}$ and the carrier transport towards the channel ($pn$-junction) with probability $T_K$ (see Fig. 7). When considering a graphene channel of width W, the conductance of M/G contact can be expressed in terms of the number of conduction modes in graphene $M_G$ and of the total carrier transport probability $T$ following the Landauer approach [48] as $G = 4q^2 T M_G h^{-1}$, with $M_G$ in the graphene channel and under the contact depending on the difference between the Fermi level and the Dirac point in the channel or under the contact respectively.

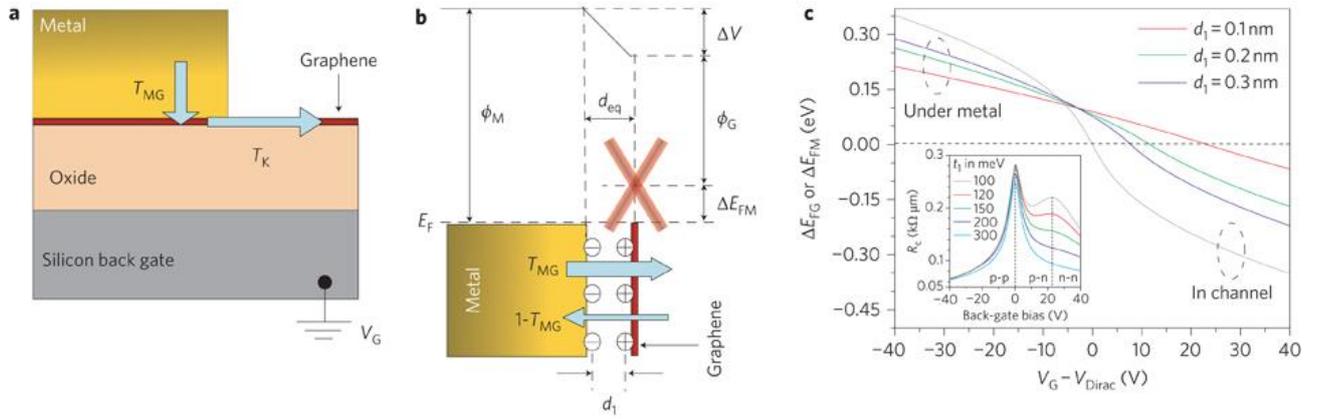

**Fig. 7.** (a) Schematic representation of carrier transport processes at the M/G interface. (b) Energy band diagram and dipole formation at the interface. (c) Difference between the Dirac-point and Fermi-level energies in the metal-doped graphene $\Delta E_{FM}$ and in graphene channel $\Delta E_{FG}$ are calculated as a function of gate bias. Inset: Calculated contact resistance $R_C$ as a function of gate bias. Reproduced from Ref. [98] with permission.

Experimentally, the contact resistance for Pd-graphene junctions significantly increases with temperature [98], differently from conventional semiconductors for which an opposite behavior is observed. This phenomenon is explained by considering possible modifications with temperature of the scattering mean free path $\lambda$ in the graphene under the contact as well as of the effective M/G coupling length $\lambda_m$, that characterizes the scattering process in the ballistic limit induced by the graphene coupling to metal. The transmission probability can be written as $T_{MG} = \sqrt{\lambda(\lambda + \lambda_m)^{-1}}$ and in the ballistic limit ($\lambda \gg \lambda_m$) it is $T_{MG} \approx 1$ while in the diffusive limit ($\lambda \ll \lambda_m$) it is $T_{MG} \approx \sqrt{\lambda/\lambda_m}$. Within this model, the effective transfer length is $L_T = \sqrt{\lambda \cdot \lambda_m}$, a value that can be very different from what generally obtained in M/S contacts, and could explain the results, not always well understood, obtained by the conventional TLM. Moreover, according to the model, the contact resistance can be reduced either by heavily doping the graphene under the metal (that causes an increase of $M_G$) or by reducing the effective coupling length $\lambda_m$ with the metal (that increases the total transport probability $T$).

In case of multilayer graphene, it has been proposed a resistor network model [99] that takes into account the screening as well as the interlayer coupling to explain the experimental data on charge and current distribution in GFETs by varying the number of layer. The dependence of the $I_{ON}/I_{OFF}$ ratio on the graphene multilayer thickness $t$ is $I_{ON}/I_{OFF} \sim 1/t$ and it is almost one for ten layers. The screening length $\lambda_{SL}$ has been estimated 0.6 nm in good accordance with previously reported data [100], and confirming a gate control limited to the first two-three layers. The second parameter of the model, $R_{int}/R_{off}$, characterize the interlayer coupling and it is found to be $\sim 0.05$, where $R_{int}$ represents the interlayer resistance (that prevents current going deep in the graphene layers) and $R_{off}$ the resistance of a single layer in the $OFF$-state.

Evidence of the current crowding effect has been reported by photocurrent spectroscopy experiment [101], in which $L_T$ at the graphene-gold interface has been obtained considering that the (closed-circuit) photocurrent is proportional to the gradient of the electrostatic potential along the interface $I_{PC} \propto \frac{d}{dx}\left[\Phi_0 e^{-x/L_T}\right]$ that causes the separation of charge carriers optically excited. Experimental data have shown that the photocurrent exponentially decreases within the gold contact, evidencing the dependence of $L_T$ on the charge carrier density. From the exponential fitting, highest values of $L_T \sim 1.6$ µm have been obtained for both electrons (for charge density $\sim 4.8 \times 10^{12}\ cm^{-2}$) and holes (for charge density $\sim -5.3 \times 10^{12}\ cm^{-2}$).

A theoretical model of the carrier transport between the 2D graphene and the 3D metal electrodes has been developed by Chaves et al. [102,103] to quantify the intrinsic factors controlling $R_C$. The physical model is based on the Bardeen Transfer Hamiltonian (BTH) method [104,105] for the calculation of $R_{mg}$ (the resistance between the metal and the graphene underneath) and on the Landauer approach [106] for the calculation of $R_{gg}$ (the resistance due to the potential step across the junction formed between the graphene under the metal and the graphene channel), where the total contact resistance is calculated as $R_C = R_{mg} + R_{gg}$. Depending on the metal and the chemical doping of channel, the two components can also result very different.

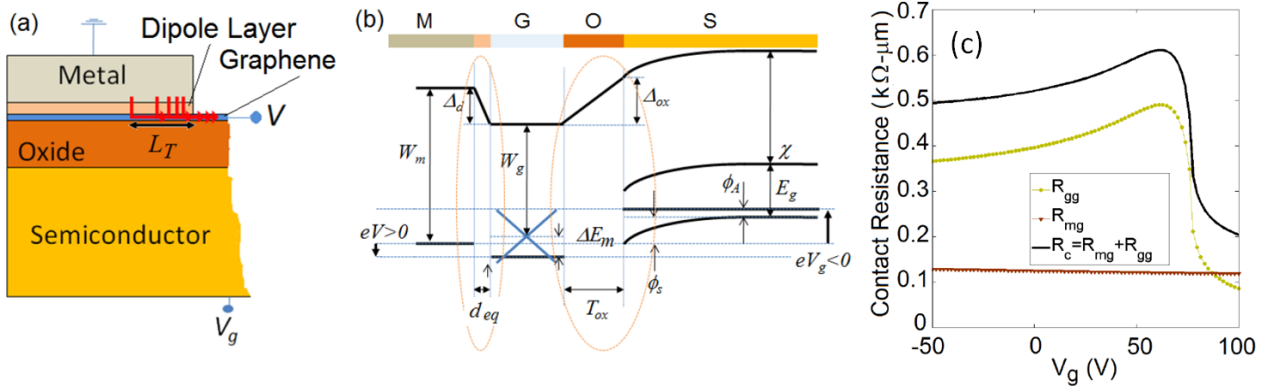

**Fig. 8.** (a) Schematic representation of the device discussed in the theoretical model by Chaves et al. [103] and (b) energy band diagram of the heterostructure metal/ graphene/oxide/semiconductor. Reproduced from Ref. [103] with permission. (c) Calculation of $R_C$ and its components $R_{mg}$ and $R_{gg}$ as function of gate bias.

The electrostatic problem is analyzed assuming that the standard back-gate GFET configuration (Fig. 8(a)) can be seen as two 1-D heterostructures, namely the metal/graphene/oxide/semiconductor in the contact region and the graphene/oxide/semiconductor in the contact region. In Fig. 8(b) the band diagram for the contact region is showed. The gate voltage dependence of the Fermi level shift in graphene under the metal ($\Delta E_m$) and in the channel ($\Delta E_g$) is modeled in the two regions respectively, these two quantities being key factors to calculate $R_C$. Indeed, according to the transmission line method [88] the component $R_{mg}$ is shown to be expressed as $R_{mg} = \sqrt{\rho_c R_{sh}^m} coth(L_c/L_T)/W_c$ with the specific contact resistivity $\rho_c$ dependent on ($\Delta E_m$), $R_{sh}^m$ the sheet resistance of graphene under the metal and $L_c, W_c$ the contact dimensions. The analytical expression of $\rho_c$ can be calculated within the BTH framework and it shows that the maximum value depends exponentially on $d_{eq}$ (equilibrium separation distance) and it is obtained for $\Delta E_m = 0$. It is also found that $R_{gg}$ strictly depends on the effective length $\lambda_{eff}$ of the potential step forming between the graphene under the metal and the channel: $R_{gg}^{-1}(\Delta E_m, \Delta E_g) = 2\pi^{-1}e^2h^{-1}W_c \int_{-k_F}^{k_F} T_{step}dk_y$ with $k_F = min(|\Delta E_m|, |\Delta E_g|)/\hbar v_F$ and $T_{step}$ the transmission probability depending on $\lambda_{eff}$. By applying the model to different metals it comes out that for Ni and Ti electrodes the $R_{gg}$ is the dominant component, while for Pd there is a competition between the two components. The breakdown of $R_C$ in its two components for Ti contacted GFET is reported in Fig. 8(c).

## 4. Improving contact resistance

Conventional contacts in graphene based devices are realized by depositing metal electrodes on top of the graphene surface. Since the beginning of the study of M/G interface, it appeared clear that this procedure often results in very large contact resistance. Several attempts were initially devoted to check different metals with proper workfunction in order to modify the Fermi-level difference between metal and graphene and to improve M/G interface. One of the best results has been obtained for Pd contacts with a contact resistance of ~200 $\Omega\mu m$ at carrier concentrations of $10^{13}cm^{-2}$ [98,107]. Several works to increase DOS in graphene and to decrease contact resistance by using different metals, surface treatments or innovative device architecture have been reported [108-115]. Some results are summarized in Table 1.

| Contacting Material | Stack layers (thickness in nm) | Deposition technique | Graphene | Measurement technique | Contact resistivity ($\Omega\mu m$) | Notes | Reference |
|---|---|---|---|---|---|---|---|
| Ag | Ag | | CVD | TLM | 1400 | *Rapid thermal annealing* | [116] |
| | Ag/Au (100/10) | E-Beam | Exfoliated | TLM | 2000 | | [113] |
| Au - Gold | Au | | CVD | TLM | 630 | *Rapid thermal annealing* | [116] |
| | Au (20) | | CVD | 2p/4p | 340 | | [101] |
| | Au | E-Beam | CVD | TLM | 1200 | *Metal on Bottom* | [118] |
| | Au (20) grains | Th. Evaporator | Exfoliated | TLM | 130 | *Grains on whole surface* | [117] |
| | Au (250) | Evaporator | CVD | TLM | 456 | *Patterned holes in graphene/edge contact* | [119] |
| | Au (81) | Evaporator | CVD | TLM | 500 | | [120] |
| | Au/Cu/Au (20/200/60) | Th. Evaporator | Exfoliated | TLM | 50 | *Resist free fabrication process* | [121] |
| Co | Co/Au (100/10) | E-Beam | Exfoliated | TLM | 300 | | [113] |
| Cr - Chromium | Cr/Au (10/20) | Th. Evaporator | Exfoliated | CBK | $10^3 \div 10^6$ | | [110] |
| | Cr/Au (100/10) | E-Beam | Exfoliated | TLM | 3000 | | [113] |
| | Cr/Au (5/150) | Sputtering | Exfoliated | 2p/4p | 5000 | | [122] |
| | Cr/Pd (0.5/40) | Evaporator | Exfoliated | HTA | 350÷750 | | [123] |
| | Cr/Pd/Au (1/15/50) | E-Beam | CVD | TLM | 270 | *Pre-plasma treatment/edge contact* | [124] |
| | Cr/Pd/Au (1/15/60) | E-Beam | Exfoliated | TLM | 100 | *Edge contact to encapsulated graphene in BN* | [125] |
| Cu - Copper | Cu | | CVD | TLM | 8800 | | [116] |
| | Cu | | CVD | TLM | 2900 | *Rapid thermal annealing* | [116] |
| | Cu (35) | Th. Evaporator | Exfoliated | TLM | 1160 | *As prepared* | [126] |
| | Cu (35) | Th. Evaporator | Exfoliated | TLM | 620 | *Annealed at 260℃* | [126] |
| | Cu (50) | E-Beam | 6H-SiC | 2p/4p | 125 | *Cuts patterned* | [127] |
| | Cu/Au (5/50) | E-Beam | CVD | TLM | 92 | *Doping by PVP/PMF insulator* | [128] |
| Fe | Fe/Au (100/10) | E-Beam | Exfoliated | TLM | 2000 | | [113] |
| Nb | Nb/Au (15/25) | Sputtering | Exfoliated | TLM | $1.9 \cdot 10^4$ | | [129] |
| | Nb/Au (25/75) | Sputtering | Exfoliated | 2p/4p | $2.4 \cdot 10^4$ | | [130] |
| **Ni - Nichel** | Ni (100) | Th. Evaporator | Exfoliated | 2p/4p | 100 | | [131] |
| | Ni (25) | Th. Evaporator | Exfoliated | CBK | 500 | | [110] |
| | Ni (60) | E-Beam | Exfoliated | TLM | 2500 | | [132] |
| | Ni (75) | Evaporator | CVD | TLM | 2200 | | [120] |
| | Ni/Au (100/10) | E-Beam | Exfoliated | TLM | 300 | | [113] |
| | Ni/Au (25/50) | Evaporator | CVD | TLM | 400 | | [120] |
| | Ni/Au (30/20) | E-Beam | CVD | 2p/4p | 2100 | | [133] |
| | Ni/Au(70/50) | Sputtering | Exfoliated | TLM | 7000 | | [134] |
| | Ni/Cu/Au (20/200/60) | Th. Evaporator | Exfoliated | 2p/4p | 1000 | | [121] |
| **Pd - Palladium** | Pd | | CVD | TLM | 570 | *Rapid thermal annealing* | [116] |
| | Pd (50) | E-Beam | 6H-SiC | 2p/4p | 457 | *Cuts patterned* | [127] |
| | Pd (75) | Evaporator | CVD | TLM | 970 | | [120] |
| | Pd/Au (100/10) | E-Beam | Exfoliated | TLM | 600 | | [113] |
| | Pd/Au (20/30) | Th. Evaporator | CVD | 2p/4p | 200÷400 | *Antidote arrays under metal electrode* | [135] |
| | Pd/Au (20/60) | E-Beam | CVD | 2p/4p | 88 | *Laser cleaning of contact area* | [136] |
| | Pd/Au (25/25) | | Exfoliated | TLM | 230÷900 | *Bias dependent* | [98] |
| | Pd/Au (30/50) | E-Beam | Exfoliated | TLM | 69 | | [137] |
| | Pd/Au (5/50) | E-Beam | CVD | TLM | 122 | *Doping by PVP/PMF insulator* | [128] |
| **Pt** | Pt/Au (25/50) | Evaporator | CVD | TLM | 1100 | | [120] |
| **Ti - Titanium** | Ti/Al (10/70) | E-Beam | Exfoliated | 2p/4p | <250 | | [138] |
| | Ti/Au (10/20) | Th. Evaporator | Exfoliated | CBK | $10^3 \div 10^6$ | | [110] |
| | Ti/Au (10/25) | E-Beam | Exfoliated | TLM | 600÷1000 | | [109] |
| | Ti/Au (10/40) | Evaporator | Exfoliated | 2p/4p | < 400 | | [139] |
| | Ti/Au (100/10) | E-Beam | Exfoliated | TLM | 800 | | [113] |
| | Ti/Au (20/80) | E-Beam | CVD | TLM | 568 | *UV-ozone treatment* | [140] |
| | Ti/Au (5/50) | E-Beam | CVD | 2p/4p | 7500 | | [133] |
| | Ti/Au (5/50) | E-Beam | CVD | TLM | 23 | *Doping by PVP/PMF insulator* | [128] |
| | Ti/Au (9/80) | E-Beam | Exfoliated | 2p/4p | 2000 | | [141] |
| | Ti/Au (9/80) | Sputtering | Exfoliated | 2p/4p | $10^4$ | | [141] |
| | Ti/Au(70/70) | Sputtering | Exfoliated | TLM | $3 \cdot 10^4$ | | [134] |
| | Ti/Pd/Au (0.5/20/30) | E-Beam | CVD | 2p/4p | 750 | | [133] |
| | Ti/Pd/Au (0.5/30/30) | E-Beam | CVD | 2p/4p | ~ 320 | *Double contact* | [142] |
| | Ti/Pd/Au (0.5/30/30) | E-Beam | CVD | 2p/4p | ~ 525 | *Top contact* | [142] |
| | Ti/Pd/Au (0.5/30/30) | E-Beam | CVD | 2p/4p | ~ 715 | *Bottom contact* | [142] |
| | Ti/Pd/Au (1.5/45/15) | E-Beam | CVD | 2p/4p | 200÷500 | *Al cap layer* | [143] |
| | Ti/Pd/Au (1.5/45/15) | E-Beam | CVD | 2p/4p | 2000÷2500 | | [143] |
| | Ti/Pt/Au | E-Beam | 6H-SiC | TLM | 20÷80 | | [115] |

**Table 1.** Summary of experiments reporting the contact resistance/resistivity when contacting the graphene with different metals. Table is ordered by material: in the first column is listed the metal contacting the graphene layer. Second column indicates the complete metal stack with thickness in nanometer. Third column reports the metallization technique (thermal evaporator, electron beam evaporator, sputtering). Then the production of graphene is specified if exfoliated, CVD, or epitaxial growth on 6H-SiC. The measurement technique indicates if $R_C$ has been measured by Transfer Length Method, two or four probe, Cross Bridge Kelvin, or Heat Transfer Analysis. Then, the reported value of $\rho_C$ is listed. In the column notes, any particular treatment or geometry discussed in the experiment is highlighted. Finally the reference from which data is extracted is cited.

The large contact resistance with graphene arises from the lack of surface bonding sites, that causes lack of chemical bonding and strong orbital hybridization [89, 127, 144-147]. Several different approaches have been exploited to reduce the contact resistance, such as work function engineering [148], cleaning of source/drain contact areas before the metallization [111,140,143], double contacts geometry [142,149], patterning of contact region [127], carbide formation [150], graphitic contact formation [151].

Nowadays, it is generally accepted that $R_C$ is determined by the chemical bonds, the electronic structures and the geometry of the interface, including both the surface and edges of the graphene. Consequently, in order to have efficient current injection at the M/G interface it is necessary to take into account the differences in the surface and edge contacts. In general, $R_C$ can be decreased by modifying the fabrication process, by performing pre- and post-treatment of graphene, or by realizing edge contacts. In the following, we will review how these possible solutions have been exploited. The achieved $R_C$ values still vary over a rather wide range and are summarized in Table 1.

*4.1. Refined fabrication process*

A way to improve the contact resistance is using a suitable device architecture. With respect the standard configuration, in which the metal is deposited on top of graphene layer, it has been shown that by realizing a "metal-on-bottom" configuration [118], the contact resistance may be improved. The standard configuration "metal-on-top" has its origin related to first available graphene layers that where mechanically exfoliated and of few micron size placed on $SiO_2$/Si substrates. The devices were successively formed on top by lithographic process often causing contamination with trapped photo-resist residues in between graphene and metal electrodes and originating large variations in contact resistance. On the other hand, the availability of CVD graphene allows the placement of graphene layers on pre-arranged structures opening also the opportunity to explore metal-on-bottom contact architectures (Fig. 9).

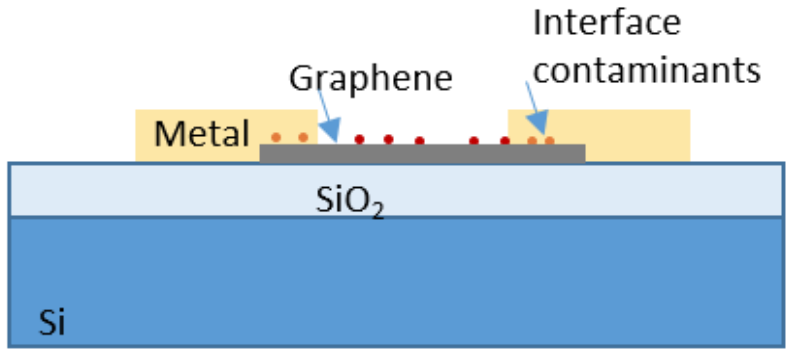

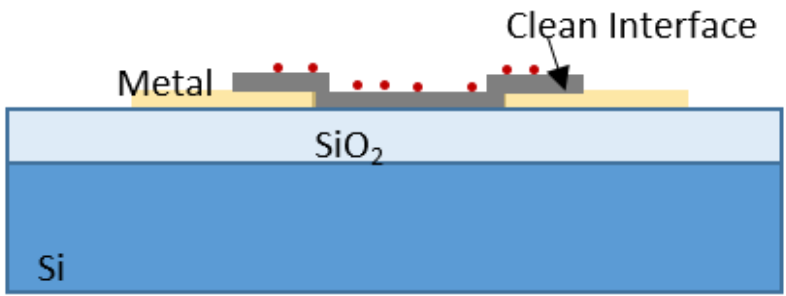

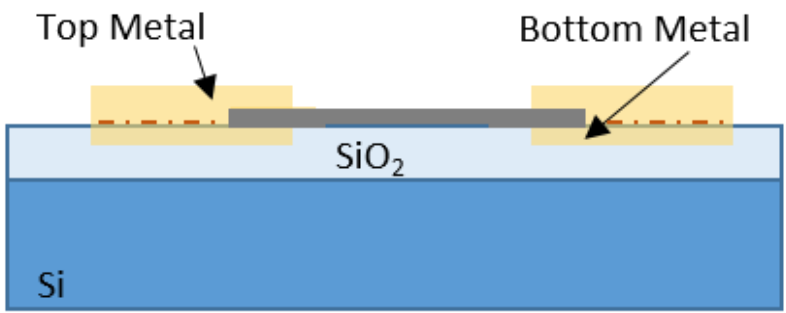

**Fig. 9.** Comparison of (a) Metal-on-Top and (b) Metal-on-Bottom configuration to contact graphene layer. A clean M/G interface is obtained for Metal-on-Bottom fabrication process. (c) Scheme of double contact configuration in which graphene flake is sandwiched between two metallic layers. Bottom contacts can be embedded in the oxide.

Typical TLM structures using different metals (Au, Pt, Pd) have been characterized in both top and bottom configuration in order to compare the contact resistance. The fabrication process for bottom contacted devices clearly provides a cleaner metal-graphene interface, the graphene surface (to be contacted) never being covered by polymer at any stage of the process. Experimental data clearly show an improved contact resistance when the bottom configuration is used. Interestingly, it is reported that also growth defects may reduce the contact resistivity by a factor of two [118]. Indeed, the contact resistivity for Pt, Au and Pd, obtained at carrier concentration varying from −1.5 to +1.5 × $10^{12}$ cm$^{-2}$, have been measured for two different defect densities, showing that for each metal, the graphene with higher defect density has a lower resistivity. Different defect density were obtained by varying the CH$_4$ gas flow rates (during graphene growth) that influences the graphene grain size and the total length of grain boundaries in the film.

The idea is to improve the adhesion of graphene on metal by modifying the fabrication method has been pursued for instance by Franklin et al. [142] who proposed a double contact geometry to considerably reduce $R_C$. In this configuration, metal electrodes are realized both below and above the graphene layer in the source/drain contact area. More precisely, first bottom contacts are realized on the substrate patterning a trench in the SiO$_2$ successively filled by e-beam evaporated Ti(5nm)/Pd(25nm) obtaining flat surface within 1nm with respect the SiO$_2$ surface. Only after transferring the CVD single layer (or exfoliated bi-layer) graphene, the Ti(0.5nm)/Pd(30nm)/Au(30nm) top contacts are evaporated completing the double contact sandwich. From the extensive characterization performed in the four-probe configuration on more than 60 samples, it has been demonstrated that the contact resistance is systematically reduced of at least 40% down to a minimum of 260 Ωμm for the single layer graphene. The higher contact resistance in bottom contacted devices is explained in terms of lower doping level in the source/drain areas with respect to the top-contacted devices. In the double contact configuration, a higher metal-induced doping may originate from the increased coverage of metal on graphene (with respect one-side geometry) favoring lower contact resistances.

Palacios et al. [143] used a 5 nm thick Al sacrificial layer between graphene and photoresist before proceeding with metal contacts fabrication. This procedure allows to reduce the graphene surface roughness from 1.2 nm, normally observed for standard processed graphene layers, to 0.23 nm, a value comparable or even better than the 0.25 nm value usually measured for as transferred CVD graphene on SiO$_2$. The increase of surface roughness is caused by photoresist residues spread on the surface. The Al cap layer is deposited on graphene soon after it is transferred on the SiO$_2$ substrate, to protect the surface. During the electron beam lithography process to realize the Ti(1.5nm)/Pd(45nm)/Au(15nm) metal stack to contact the device, only the source drain areas are etched with a developer containing tetra-methyl ammonium hydroxide to remove Al cap layer and expose pristine graphene. The electrical characterization demonstrates an improvement of the contact resistivity from 5 to 10 times with respect twin sample without Al layer, the lowest values measured in the range 200-500 Ωμm.

*4.2. Surface treatments*

To fabricate GFETs with CVD grown graphene it is necessary to transfer the monolayer from a metal foil to a substrate, typically using a polymer Poly(methyl methacrylate) (PMMA). Such polymer is also used as standard photoresist for lithographic processes on graphene by EBL. Usual cleaning by acetone leaves some residues on graphene surface, mostly due to the van der Waals interactions as well as chemical bonds [152]. The residues cause the shift of the Fermi level, the reduction of carrier mobility [130,153,154] as well as the increase of contact resistance [121,155] due to the contaminants at the M/G interface. In order to reduce contact resistance, several methodologies of surface treatments have been proposed.

Important reduction of contact resistance down to 200 $\Omega\mu m$ (for Ti/Au electrodes) has been reported as result of ultraviolet/ozone treatment on CVD graphene monolayer, while preserving the electrical properties [156]. This treatment is commonly used in semiconductor technology for surface cleaning and it is performed during the lithographic process before deposition of metallic contact pads, soon after opening the window in the photoresist masking layer. By atomic force microscopy and Raman spectroscopy it has been demonstrated an almost complete removal of contaminants (coming from transfer as well as photolithography process) from graphene surface for exposure time less than 30 minutes, the improvement of contact resistance starting already after 10 minutes treatment. X-ray photoelectron spectroscopy allowed a quantitative analysis of the chemical and disorder changes induced by the ultraviolet/ozone treatment [140]: the signal related to carbon increases after the lithographic process probably due to the photoresist contamination of the surface, as confirmed also from the appearance of the fluorine signal. The ultraviolet/ozone treatment reestablish the carbon percentage as was before the lithographic process. Correlation of the contact resistance with the surface treatment has been performed by massive data acquisition in TLM configuration to extract the contact resistance from more than one hundred samples. The range of variability of $R_C$ values resulted much smaller than what observed for untreated devices, due to the uncontrollability of the surface contaminants. For 25 minutes treatment, an average value of about $500\pm200$ $\Omega\mu m$ is reproducibly obtained with respect the values of $40\div45$ $k\Omega\mu m$ obtained for unexposed devices.

Since 2007 it was reported that oxygen plasma treatment of graphene FETs could possibly increase the bonding strength despite the dangling bond generated by the treatment may significantly degrade mobility [157]. An extensive study has been performed [158] on 250 back-gated graphene FETs, fabricated by transferring CVD graphene on standard Si/SiO$_2$ (300 nm) substrate. The plasma treatment was limited to the electrode areas before the metal (Pd, 50 nm) deposition, and various gases (H$_2$, N$_2$, Ar, and O$_2$) were tested, reporting enhanced adhesion between graphene and metals when $O_2$ was employed. Different techniques were used to identify the effects on the graphene surface: contact angle measurement for the wettability, Raman spectroscopy for the defects and bonds, Fourier-transformed infrared spectroscopy for the functional groups. Finally, electrical measurements reported hole mobility up to 5200 $cm^2V^{-1}s^{-1}$ after 4 seconds plasma treatment, suggesting an important increase of the surface adhesion that could allow the possibility to avoid the process step of a buffer layer in fabricating GFETs.

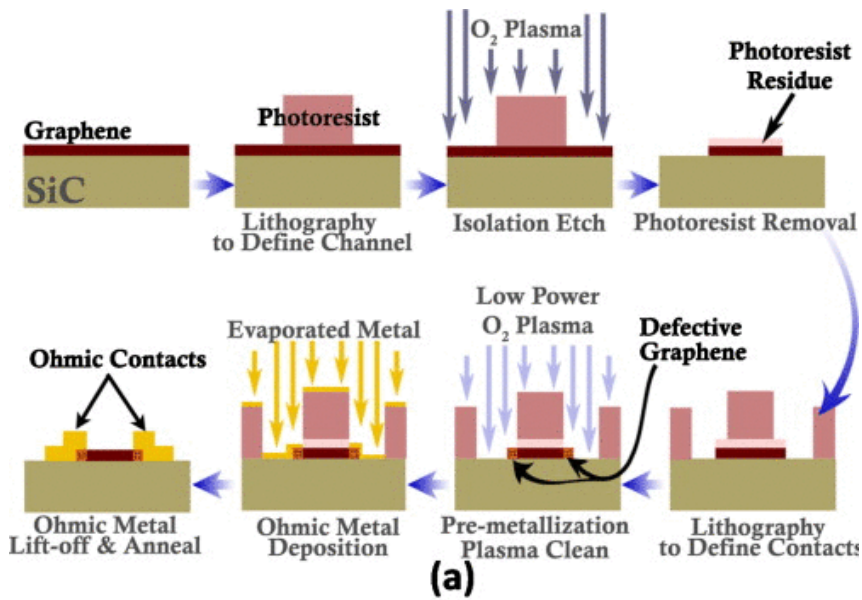

(a)

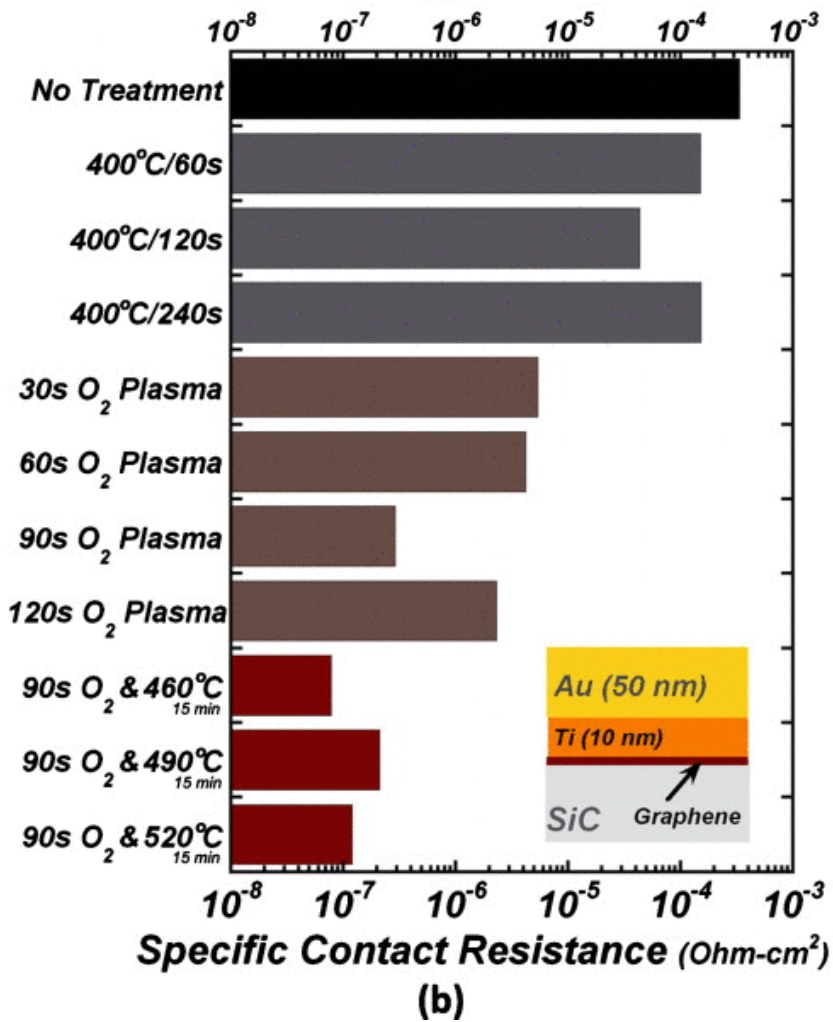

Specific Contact Resistance (Ohm-cm²)

(b)

**Fig. 10.** (a) Scheme of the fabrication process in which O₂ plasma treatment is performed before metal evaporation of contacts. (b) Effect of surface treatments on the contact resistance for Ti/Au metal stack on graphene. Figure reproduced from [111] with permission.

Low power $O_2$ plasma treatment has been demonstrated to realize high quality ohmic contacts to epitaxial graphene, grown on SiC [111]. The treatment was performed just before the metal deposition, and it was followed by furnace annealing in nitrogen gas (Fig. 10(a)). Interestingly, this method allows reproducible contact resistance for Ti/Au contacts below $10^{-7}\Omega cm^2$, to be compared to the standard values $> 10^{-5}\Omega cm^2$ obtained when lithographic processes are used without any treatment to improve the metal/graphene interface quality. Treatments up to 90 seconds have been demonstrated to be effective to improve the contact resistance despite the increase of defects in graphene. Indeed, graphene and SiC peaks appear reduced in X-ray photoelectron spectroscopy performed after the lithographic process probably due to the presence of resist residuals; the contaminants are removed after $O_2$ plasma treatment. Besides, plasma treatments longer than 90 seconds have been demonstrated by Raman spectroscopy to cause degradation of graphene. However, the degradation is limited to the graphene area that has to be covered by the metal, the channel remaining unaffected. The best contact resistance ($\sim 7.5 \times 10^{-8}\Omega cm^2$) has been obtained for 90 seconds $O_2$ treatment followed by 15 minutes thermal annealing at temperature between 450-475 °C (Fig. 10(b)).

The application of low-power $O_2$ or ultraviolet ozone to clean the graphene, just discussed above, favors the surface modification towards a hydrophilic nature as well as creating defects in the contacts region before the metallization process, in order to improve M/G interaction through chemical bond formation. The principal drawback of such approaches is the random introduction of defects that could cause important scattering in the contact area, detrimental for the contact resistance, if fine tuning of the treatment process is not developed.

Recently, a laser cleaning treatment of the graphene surface to remove polymer (PMMA) residues has been characterized in detail [159]. The technique has been applied to exfoliated graphene as well as CVD graphene films, both types placed on standard Si/SiO$_2$ substrates, in order to produce high quality GFETs. The study has been systematically performed on graphene films with different number of layers, for various laser exposure powers and times, finding as the best laser cleaning conditions for monolayer graphene 30 mW power for 180 s. Separated experiments were then performed to analyze the effects of such laser treatment on channel and contact areas. The electrical characterization of Ti/Au-contacted GFETs, in which only the channel was treated, has been performed by measuring transfer characteristics in the standard two-probe setup: comparing the mobility obtained for as-fabricated devices ($\mu_e = 2141$ cm$^2$V$^{-1}$s$^{-1}$; $\mu_h = 2230$ cm$^2$V$^{-1}$s$^{-1}$) with the values measured for the treated devices ($\mu_e = 3770$ cm$^2$V$^{-1}$s$^{-1}$; $\mu_h = 4232$ cm$^2$V$^{-1}$s$^{-1}$), it has been demonstrated that the cleaning process on the channel is effective, causing increased mobilities of a factor 1.5-2.6. The same laser treatment has been tested on the contact areas before the metal (Pd/Au) deposition on CVD monolayer graphene to realize the GFET devices. A systematic reduction of the contact resistivity has been obtained with respect to untreated devices, with an average contact resistivity of 107 $\Omega\mu m$ (for treated devices), well below the previously reported values in the range 150-185 $\Omega\mu m$ for Pd contacts [98,160].

Despite the good results discussed above, the most common procedure to remove resist residue (introduced during the fabrication process) from the contact areas is the thermal annealing [110, 121, 161, 162]. Above 200°C the decomposition of the resist residues takes place [152, 161], but if annealing is performed after the metal deposition of the contact electrodes, it is difficult that residues may be removed. Indeed, for Ni-contacted graphene it has been reported that there is no relevant improvement of contact resistance after thermal annealing performed on the already contacted device [162].

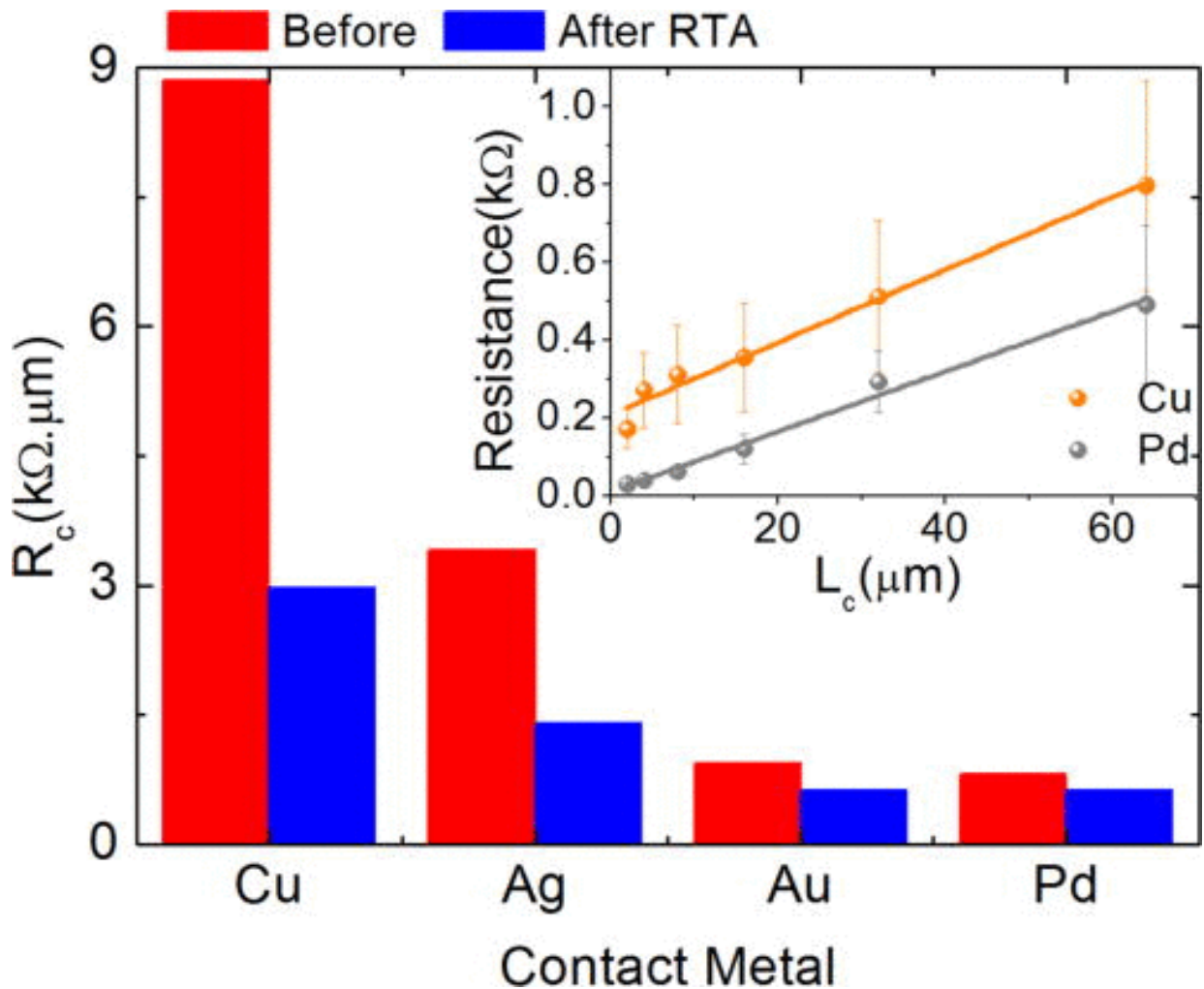

**Fig. 11.** Contact resistance in graphene transistors for different contact metals before and after rapid thermal annealing (RTA). The inset shows the total resistance vs channel length for Cu and Pd contacts. Figure reproduced from [116] with permission.

Thermal annealing treatments have also been reported as effective process to obtain $R_C$ lowering [111, 116, 126]. TLM measurements of $R_C$ on Cu-contacted graphene [126] showed that very low contact resistance could be achieved after long annealing time (above 12 h) at about 300°C in high vacuum (about $10^{-7}$ torr) with reported values going from 2 k$\Omega\mu$m$^2$ (for as prepared samples) to 0.1 k$\Omega\mu$m$^2$ (for annealed samples). Rapid thermal annealing of graphene/metal contact has been reported for various metals (Cu, Ag, Au, Pd) [116]. The treatment was performed by means of halogen light furnace on GFETs realized by CVD graphene transferred on Si/SiO$_2$(100nm) substrates. The annealing process was repeated several times (up to ten) controlling either the ramp up (from room temperature to 300°C in 40 s) than the ramp down (to room temperature in 5 minutes), keeping the annealing temperature fixed at 300°C for 1 minute. Measuring the transfer and output characteristics of the GFETs, it was demonstrated that such rapid thermal annealing was effective in reducing the contact resistance for the various metals, as summarized in Fig. 11.

Very interestingly, Leong et al. [163] have reported a detailed study performed on Ni-contacted back-gated GFETs to compare the effect of thermal annealing on devices with resist-patterned contacts and with resist-free contacts. All devices were electrically characterized in high vacuum by standard four-probe measurement method to extract the contact resistance. The observation of a similar improvement of the average $R_C$ due to the thermal annealing for both kind of devices (with and without resist) is considered as an indication that the principal effect is not related to the removal of residues. They argue that the improvement of contact resistance is mostly due to the dissolution of graphene atoms into metal contact at the Ni/graphene or Co/graphene (chemisorbed) interfaces: the chemical reaction at the interface causes the formation of chemical metal/graphene-edges bonds, originating the contact resistance reduction. Such interpretation suggests a possible route for further reduction of the contact resistance is to maximize end-contact geometry between metal and graphene. Several other methods have been reported in literature concerning the improvement of electronic properties of graphene by surface treatments. In the following, we just mention some relevant studies because they focalize mostly on the carrier mobility, without measuring the impact on the contact resistance. A wet-chemical approach has been proposed by Cheng et al. [164] using chloroform as solvent for resist residues. This method prevents the heavy doping and degradation of mobility in graphene that could be induced by the simple thermal annealing at high temperature due to the interaction with the SiO$_2$ substrate. Another wet-chemical process is the exposure of the device to formamide, proposed by Suk et al. [165]. In this case, a shift of the Dirac point (in $p$-doped graphene) towards zero gate voltage and an important increase of carrier mobility (of about 50%) at room temperature is explained in terms of electron donation to graphene by the −NH$_2$ functional group in formamide.

Current induced annealing has been also demonstrated to improve the electronic properties of graphene [122, 166] by applying high current trough the flake causing several mW of power dissipation that is able to evaporate adsorbates from the surface without degrading the graphene. The enhancement of carrier mobility due to current annealing has also been demonstrated in suspended graphene, reaching values as high as $2.3 \times 10^5$ cm$^2$V$^{-1}$s$^{-1}$ [167].

### 4.3 Edge contacts

Another innovative solution is to realize a different contact configuration to graphene, named edge contact. Differently from the usual top/bottom-contact, for the edge-contact the metal electrodes are connected to graphene layer along one-dimensional edge (see Fig. 12) [124, 128, 131, 145, 168-170].

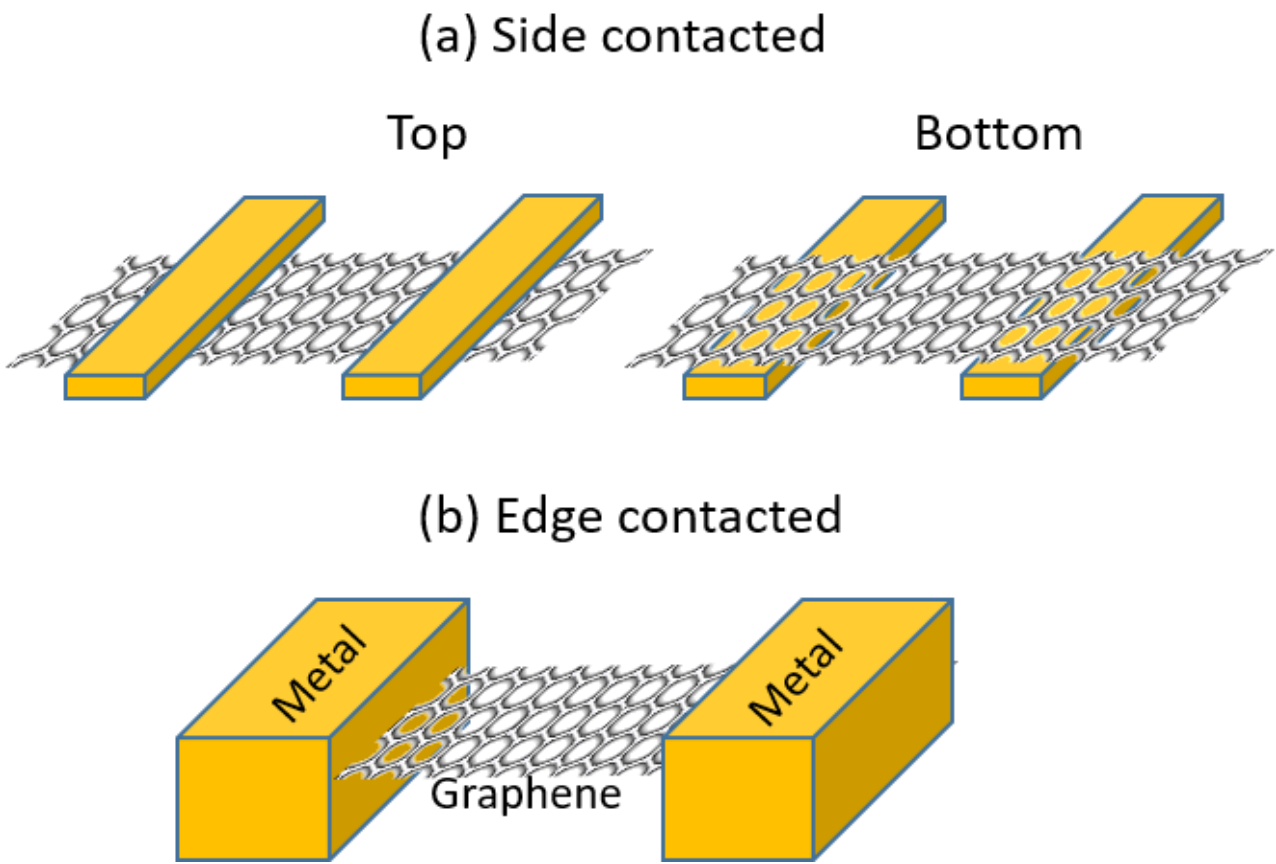

**Fig. 12.** Scheme of the metal-graphene contact configurations. (a) Side contacted devices can be realized either as top- or bottom-contacts. (b) Edge contacted device also named end-contact configuration.

Matsuda et al. [169] theoretically analyzed the edge-contact structure realized between graphene and several metals (Ti, Pd, Pt, Cu, Au) showing that it causes an important reduction of the contact resistivity due to a higher cohesive energy at the interface between the carbon atoms and the metal. This result has been achieved by using first-principle quantum mechanical density functional and matrix Green's function methods to determine the current/voltage characteristics and the contact resistance for edge-contacted M/G interfaces. From the model, it comes out that the density of states near $E_F$ depends mostly on the C $p$-orbitals and surface metals $d$-orbitals. However, while for standard top-contacts only the carbon $p\pi$ orbitals of carbon atoms contribute to the cohesion to the surface metals, in the edge-contact case also $p\sigma$ orbitals contribute to the surface cohesion and transmission. By comparing the calculated contact resistances in the two configurations, it is demonstrated that for edge-contacted M/G interfaces the contact resistance per C atom is significantly reduced for all metals, the maximum factor (more than 6000 times) being expected for Au metal electrodes. Moreover, the smallest contact resistance per surface C atom is expected for Ti electrodes.

Despite the clear theoretical prediction of important advantages to realize edge-contacted structures to graphene, it represents a difficult task from an experimental point of view.

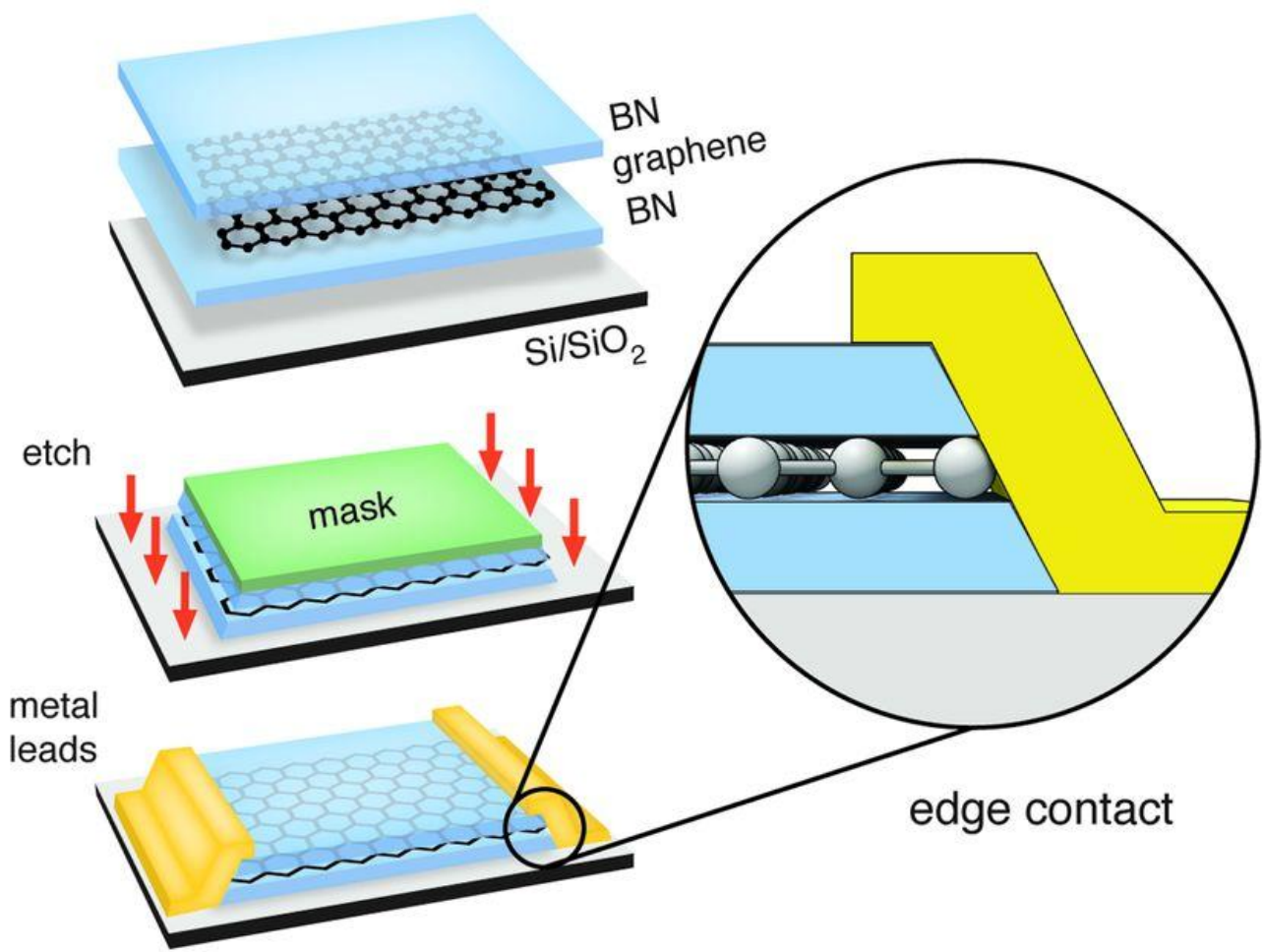

**Fig. 13**. Schematic representation of the fabrication process of edge contacts to encapsulated graphene. Figure reproduced from [125] with permission.

The fabrication of one dimensional edge contact between 3D metal and 2D graphene layer has been reported by Wang et al. [125] starting from a sandwiched structure where a graphene sheet is encapsulated in between two hexagonal boron nitride layers. The layered structure is protected on top by an hard mask and it is then plasma etched on the sides in order to obtain a free edge of the graphene layer to be metalized in a completely polymer-free fabrication process for the M/G interface (see Fig. 13). The true edge contact to graphene has been confirmed by cross section scanning transmission electron microscope image, no metal atom being diffused in the graphene/boron-nitride interface. Despite the one dimensional contact edge could affect the carrier injection, the experimental TLM data evidenced contact resistivity as low as 100 Ωμm for some devices. Moreover, such devices also showed room-temperature mobility up to $1.4 \cdot 10^5$ cm$^2$V$^{-1}$s$^{-1}$ and sheet resistivity below 40 Ω/□ at carrier density $n > 4 \times 10^{12}$ cm$^{-2}$. Interestingly, the contact resistance is found to be inversely proportional to the contact width and independent of temperature.

From a theoretical point of view, the low contact resistance observed for the edge contacted devices is explained [125] in terms of shorter bonding distance with larger orbital overlap in edge configuration with respect surface contacts, as found from Ab initio simulations in the framework of a first-principle atomistic model, in accordance with Matsuda et al. [169].

Smith et al. [127] have modified the contact regions by realizing several parallel cuts on graphene (Fig. 14), through a reproducible lithographic process, in order to maximize the length of graphene edges bonded with the metal. Hundreds of two-terminal Cu contacted devices (created on epitaxial graphene grown on semi-insulating 6H(0001) SiC substrates) have been tested, measuring the average total resistance $R_{tot}$ before and after heated vacuum annealing at T=350°C for increasing number of cuts. Moreover, TLM structures have been characterized to quantify the effect of cuts on the contact resistance.

A rapid increase of the total resistance with the number of cuts has been reported for not-annealed devices, while a reduction of the total resistance is observed to a minimum value (for eight cuts) is observed for post-annealed devices. For higher number of cuts $R_{tot}$ starts to raise again, however remaining well below the values for not-annealed case. Such behavior has been explained postulating that the carrier transmission across the M/G interface is increased by the bonding to cut edges, the annealing process enabling this phenomenon. Moreover, the increase of $R_{tot}$ for larger number of cuts produces too narrow graphene strips (below 40 nm) causing important reduction of mobility and/or a Schottky barrier detrimental for the carrier injection. The characterization of TLM structures has also given an estimation of the contact resistance, evidencing that the reduction due to the annealing is of about 80% for patterned contacts (125 Ωμm being the best value obtained), in comparison with a limited annealing induced reduction of 30% in case of no-patterning. The same study [127] also verified the effects of patterned contacts on Pd-contacted top-gated GFETs reporting a 22% reduction of contact resistance as well as an improved device-to-device consistency, a relevant feature for potential applications.

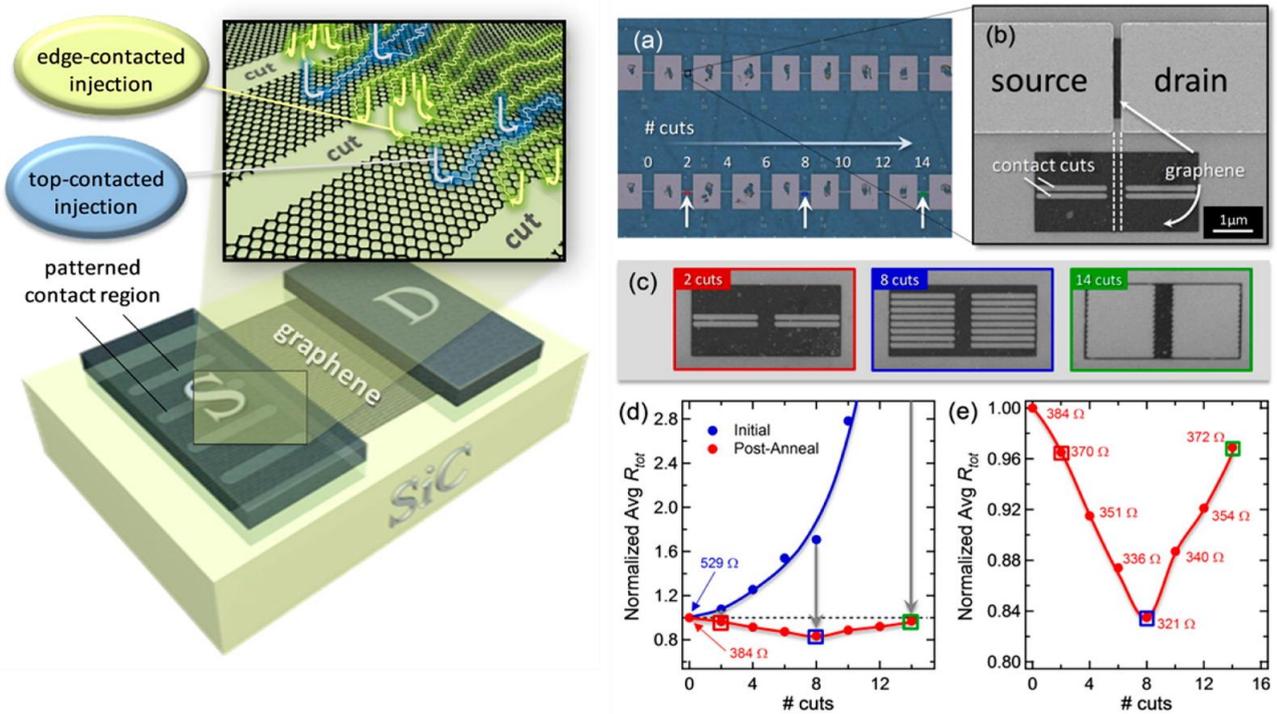

**Fig. 14**. Scheme of patterned contacts in two-terminal graphene device. Carrier injection occurs both along the cut graphene edges and from the graphene surface. (a) Optical image of the graphene devices with raising number of cuts in the contact regions. (b) SEM image of a graphene device with two cuts. (c) SEM images of devices with 2, 8, and 14 cuts. (d) Average total resistance before and after vacuum anneal. (e) Zoom-in of average total resistance after the annealing. Figures adapted with permission from [127].

Leong et al. [131] reported a different approach based on Ni-catalyzed etching process in hydrogen to produce a large number of defect-free edge contacts to graphene and to obtain very low contact resistance. By depositing Ni at the source/drain contact area and performing thermal annealing in hydrogen they realized multiple nanosized pits (see Fig. 15) with zigzag edges to favor strong chemical bonds with the successively deposited contact metal (nickel 100 nm thick). The etched pits are the result of a known Ni-catalyzed gasification process C (solid) + 2H$_2$ (gas) + CH$_4$ (gas) [171].

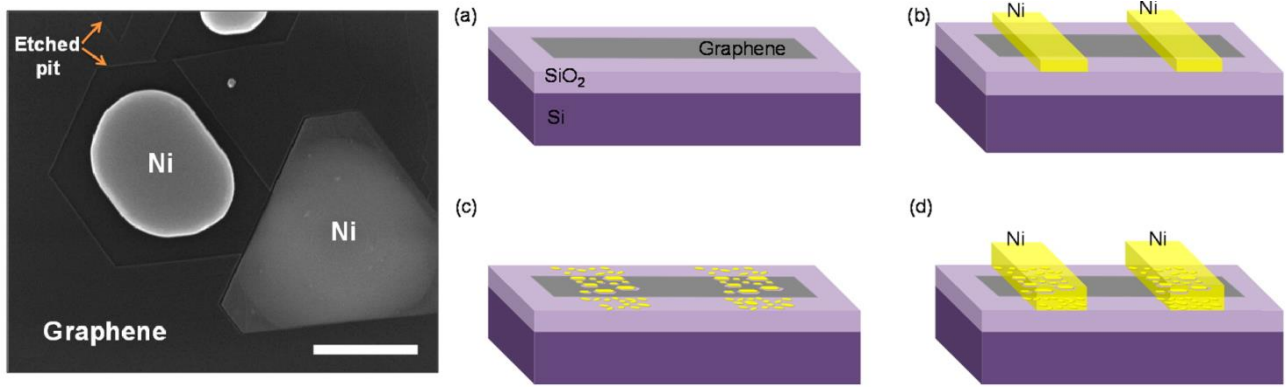

**Fig. 15.** SEM image of graphene surface after contact treatment. (a) Schematics of the fabrication process. Exfoliated graphene is placed on a Si/SiO₂ substrate and patterned into a strip. (b) Thin Ni films are deposited at the source/drain regions and successively annealed in hydrogen producing (c) large amount of pits enclosed by zigzag graphene edges. (d) Ni electrodes are finally deposited to contact the graphene device. Figures are adapted with permission from [131].

Electrical characterization by four-point probe measurement technique on several devices has demonstrated that the Ni-catalyzed etching process allows to obtain extremely low contact resistances, 89 Ωμm for single layer graphene and 11 Ωμm for bilayer graphene. These values resulted much lower than twin untreated devices. This confirms that the contact resistance is significantly affected by the amount of edge-contacts created in the planar graphene device. Moreover, the total length of formed graphene edges is expected to increase with the etching time. By characterizing devices produced with different etching time, they found that the $R_C$ reduction is effective for etching time up to ten minutes. For longer etching, no further improvements are recorded.

Park et al. [128] investigated the effects of *n*-doping in graphene in combination with edge contact configuration in order to reduce the contact resistance arising at the M/G interface, reporting a record low value of 23 Ωμm at room temperature for CVD graphene. The graphene *n*-doping is obtained by charge transfer from poly(4-vinylphenol)/poly(melamine- co -formaldehyde) (PVP/ PMF) insulator. Indeed, triazine functional groups, which are electron-rich aromatic molecules, are present in PMF and have nitrogen atoms working as electron donors [69] when interfaced with graphene. Tuning the PMF to PVP ratio is possible to modify the doping, the graphene being turned to *n*-type above 200% PMF concentration. For 400% PVP/PMF layer, large *n*-doping of graphene produces higher density of states and lower contact resistance, compared to surface contact resistance on pristine graphene. By testing different metals (Ti, Pd, Cu), it was also observed that due to the *n*-doping, the surface contacted metals did not affect the graphene Fermi level as in pristine graphene because the already larger DOS, resulting in a contact resistance less dependent on the metal type. Moreover, in order to further reduce contact resistance, edge contacts were realized by patterning contact area with various configurations (varying the ratio perimeter/area of the patterning). TLM measurements have demonstrated reduced contact resistance for all tested metals with a 1 μm pattern periodicity of the contact area, reporting 83 Ωμm for Ti, 254 Ωμm for Cu and 484 Ωμm for Pd. Reducing the patterning periodicity to 150 nm the record value of 23 Ωμm is reported for Ti edge contacts.

A different method to favor the edge contact at the metal/graphene interface in the contact region has been proposed by Passi et al. [119]. CVD graphene was used to produce Au-contacted TLM structures. Before the metallization, graphene surface was etched in order to create a regular array of holes with varying diameters from 50 to 1000 nm. By comparing the electrical characterization of devices with or without the patterned holes, it has been experimentally demonstrated a reduction of the contact resistance due to formation of edge contacts from 1518 Ωμm to 456 Ωμm for holes of 500 nm diameter.

To enhance the bonding and the coupling at the metal/graphene interface with an edge-contact configuration Yue et al. [124] have used an oxygen pre-plasma process. During the fabrication process of back-gated transistors, the CVD-graphene was exposed to $O_2$ plasma treatment for increasing durations up to 65 s, in order to cause the formation of edge contacts before the metallization of contacts. A large number of devices has been produced for TLM characterization and the statistical analysis of the results evidenced a reduction of the contact resistance of about 77% for plasma treatment duration of 45 s. It has been reported that in the range from 15 s to 45 s the contact resistance can be expressed as $R_C = -90t + R_{C0}$ where $t$ is the duration of the plasma treatment and $R_{C0}$ is the contact resistance before the treatment. The lowest achieved value is 270 Ωμm. For longer durations, the contact resistance raises again, due to the excessive damaging of the graphene. Raman spectroscopy data have been reported to confirm the formation of defects at the exposed graphene edges due to the treatment. The efficiency of the treatment is explained as a partial replacement of C-C bonding in graphene with C-O bonds and the presence of edge contacts with O termination are responsible for the important lowering of $R_C$. This interpretation is based on the knowledge that incorporation of interfacial species as oxygen at the graphene edge before the formation of the M/G contact enhances the transmission through the interface [125].

A similar approach to realize edge contacts and to lower contact resistance is reported by Song et al. [135]. They produce a regular array of antidots patterned in the contact region of CVD graphene by oxygen plasma etching in combination with electron beam lithography (Fig. 16).

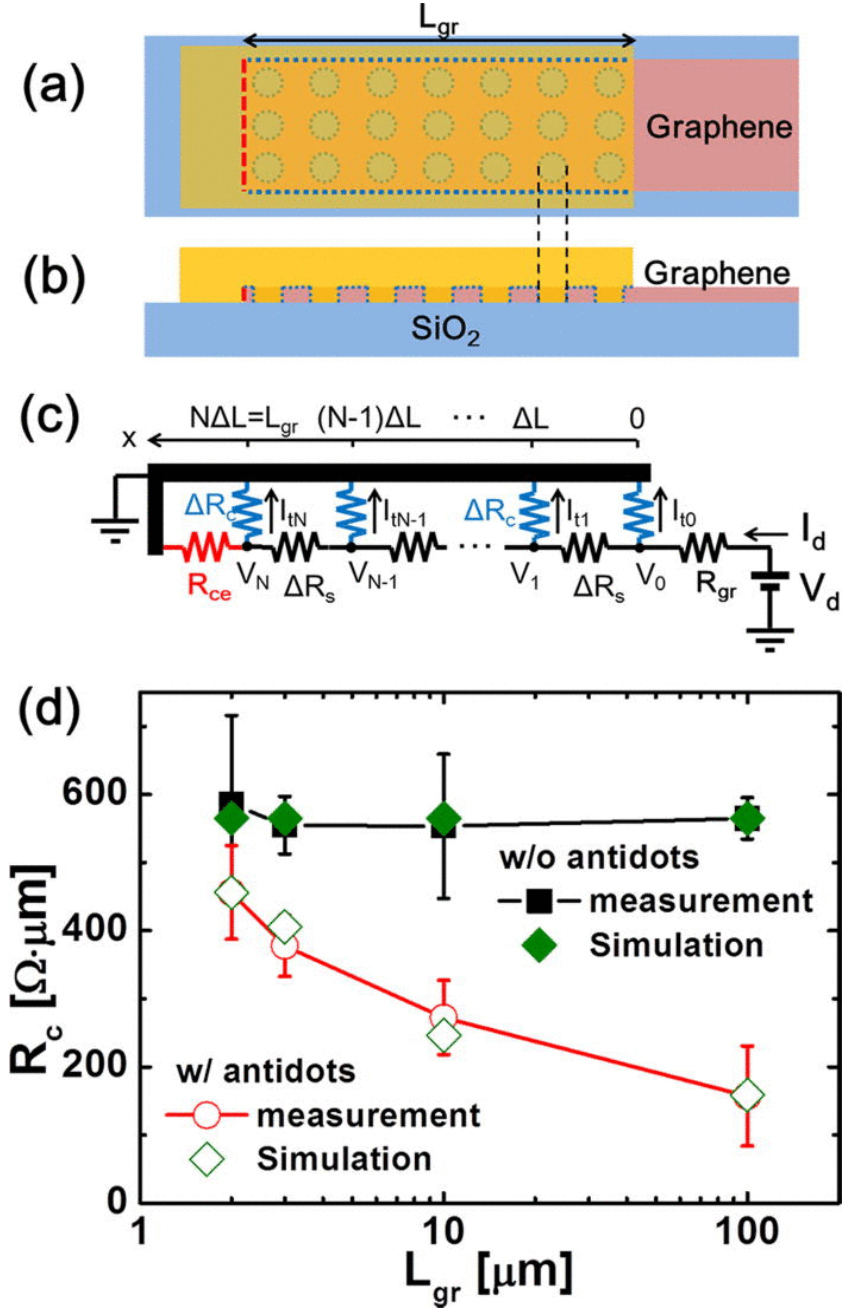

**Fig. 16.** (a) and (b) Schematics of graphene device with patterned antidots . (c) Theoretical 1D model of graphene-metal contact. (d) Experimentally extracted and simulated contact resistances with or without graphene antidots, as a function of graphene contact length. The figure is reproduced from [135] with permission.

The devices with engineered contacts show improved contact resistance as well as better carrier mobility and drain current. By modifying the antidots radius, for fixed total etched area, they verified the effect of different contact lengths $L_{gr}$ on the device performance. By comparing experimental results and theoretical simulation, the specific contact resistivity has been extracted as $2.2 \cdot 10^{-9} \ \Omega cm^2$ for the lowest $L_{gr} \approx 100 \mu m$, the resistivity increasing for shorter $L_{gr}$.

## 5. Superconducting and ferromagnetic contacts to graphene

The interface between graphene and a superconducting material (Sc) represents a unique opportunity to study the interaction of massless Dirac fermions with Cooper pairs, and opens the opportunity for the development of innovative devices. Graphene may acquire either superconducting or ferromagnetic properties by proximity effect [172-175]. Form a theoretical viewpoint, it has been demonstrated that increasing the electron–phonon coupling through the coating of alkali atoms on graphene makes it possible to induce a superconducting state [176]. Indeed, the superconducting state has been observed experimentally in metal-decorated graphene [177].

Beenakker et al. [178-180] suggested that graphene enables transparent electrical contacts with superconductors and represents an ideal system to study the physics of superconductivity at mesoscopic scale.

Heersche et al. [173] realized back-gated GFETs with Ti/Al contacts and demonstrated that a (bipolar) supercurrent, carried either by electrons in the conduction band or by holes in the valence band, can flow between two superconducting electrodes in short graphene channels.

Rickhaus et al. [181] used Nb electrodes to realize Sc–G–Sc devices with exfoliated graphene on a $SiO_2$/p-Si substrate. They studied the integer quantum Hall effect as well as Andreev processes at the G–Sc interface. Their devices were also tested as back-gated field effect transistors, showing asymmetric transfer characteristics with saturation in the p-branch and a field effect mobility around $3000 \ cm^2V^{-1}s^{-1}$. They also reported increasing contact resistance for decreasing temperature: to obtain transparent contacts they used a 4 nm Ti interface layer between Nb and graphene.

Sc–G–Sc junctions were also fabricated using Nb or ReW by Komatsu et al. [182] to study the superconducting proximity effect through graphene. They used 4 to 8 nm Pd interface layer to improve the transparency at the interface and were able to detect a reduction of the critical current near the graphene charge neutrality point, due to specular Andreev reflection.

High-quality Josephson junctions with supercurrents above 2K were fabricated by Mizuno et al. [183] using suspended monolayer graphene–niobium nitride (NbN). The quality of contacts represented the major challenge and was solved by inserting a Ti/Pd thin interlayer before Nb sputtering.

To date, Nb is the most used material for Sc/G interface studies, both for the relative high critical temperature and for the well understood properties; however, an ultra-thin interface layer is often introduced to achieve a lower contact resistance. Detailed studies of the Nb/G contact without any additional interface layer have been reported [129,130], with specific contact resistivity $\rho_c \approx 19 \ k\Omega\mu m^2$ and carrier mobility as high as $4000 \ cm^2V^{-1}s^{-1}$ for back-gated GFETs on highly doped p-Si/$SiO_2$ substrate. Those works also showed that asymmetric transfer characteristics with a resistance plateau in the n-branch can arise from the the p-doping effect of the weakly chemisorbed Nb.

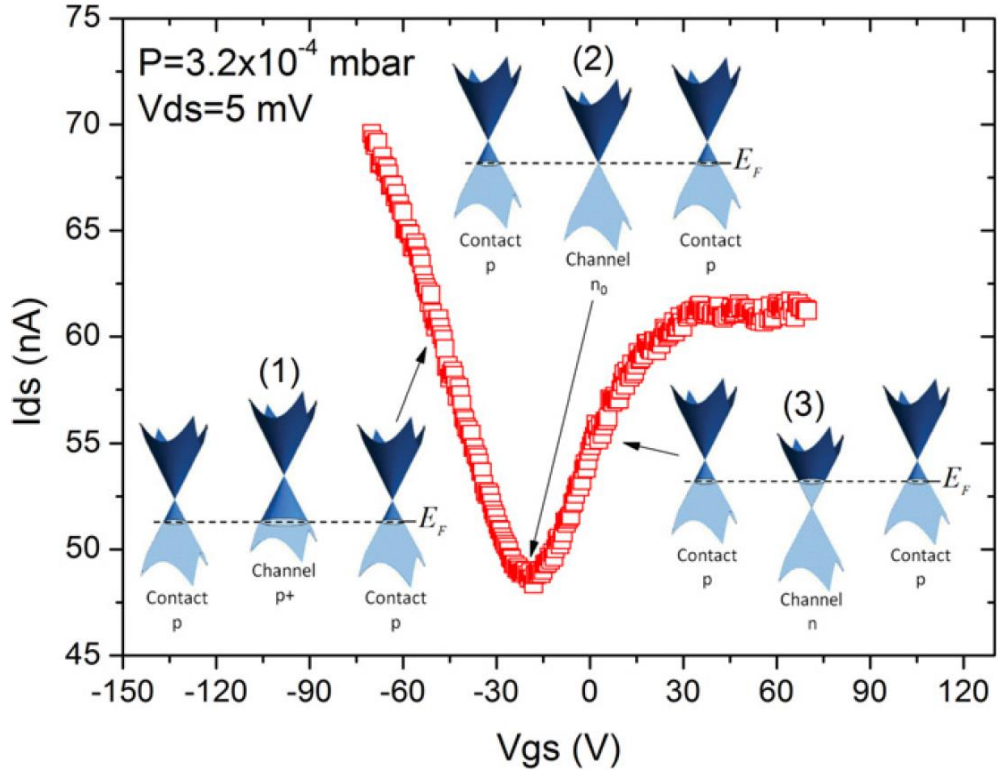

**Fig.17.** Transfer characteristics measured for a back-gated GFET on highly doped p-Si/SiO₂ substrate contacted by Nb electrodes. Fermi level for graphene at the contacts and in the channel accounting for the current behavior as a function of Vgs. It is assumed that the Fermi level is not pinned at the contacts, where p-doping occurs. The figure is reproduced from [130] with permission.

Transfer characteristics with similar asymmetry have been reported [134, 148, 122, 184-188], with feature appearing as plateau or as double dip in the n- or p-branch. Pd [98], Ni [134] or Pt [185,186], with a higher workfunction than graphene, cause p-doping and thus a lower conductance in the n-branch, while n-dopants like Cr [122] or Ti [188], originate lower conductance in the p-branch.

Likewise, ferromagnetic electrodes on graphene have attracted large interest due to the possibility to realize spin transport in graphene based spin valves [189]. Indeed, graphene represents one of the most promising candidates as material for spintronic applications due to long spin relaxation times and lengths that have been theoretically predicted [190,191]. Spintronics deals with the development of innovative solid state devices for storage and logic operations by using the spin degree of freedom of electrons [192]. The standard configuration requires ferromagnetic electrodes contacting the conduction channel. Several attempts to realize graphene based spin-transistor as well as spin valves have been reported [193-200], but the measured values of spin lifetimes always resulted significantly smaller than the prediction.

It has been identified that one of the more important factors influencing the spin lifetime is the nature and the quality of the interface between the magnetic electrode and graphene [201]. For example, tunneling contacts suppress spin relaxation, while low resistance barriers cause uncertainty in the lifetime determination.

Several reports on spin injection and transport in graphene use the non-local geometry [194,195,202-215] in which the voltage probes detecting the spin density are positioned out of the charge current loop (using other electrodes) [216]. Due to the relevance of contacts to graphene on the spintronic properties a further classification is generally done based on the nature of the interface between G and the ferromagnetic electrodes: (i) transparent contacts, where G is in direct contact with the magnetic electrode [217] or through a non-magnetic metallic layer [207]; (ii) pinhole contacts, in which the magnetic materials contact the graphene layer through holes in an insulating barrier, such as $Al_2O_3$ [218]; (iii) tunneling contacts, where a thin insulating barrier exists in between the graphene layer and the magnetic electrodes. Several solutions have been reported including h-BN [219], fluorinated graphene [214] and $Al_2O_3$ grown by atomic layer deposition [208], $TiO_2$ [202,204], MgO [203,215].

A detailed study on contact-resistance dependence of spin lifetime, measured in graphene via the Hanle spin precession technique in non-local spin valves, has been reported by Sosenko et al. [201]. They find an expression for the precession curves with finite contact resistance and analytically reproduce the different regimes arising by varying diffusion length, contact resistance and device dimension.

Cobalt nanosheets of variable thickness have been also used to realize Co/G/Co spin valves to study spin-magnetotransport in the presence of localized spins [220]. For thin cobalt sheets, a negative magnetoresistance is observed over a wide temperature range from 5K to room temperature, while a magnetoresistance sign change from negative to positive is recorded for increasing temperature when using thick Co sheets. Such phenomenon is explained in terms of spin polarization in graphene induced by the ferromagnetic nanocontacts.

Ref. [189] reports a study of the contact resistance and spin signal in CVD graphene contacted by various ferromagnetic materials (Co or CoFeB) in direct or tunnel configuration (using $Al_2O_3$ or MgO as insulating layer) for electrical spin injection and detection. By analyzing the electrical characteristics, it is found that the resistance-area product results severely scattered in a range from several k$\Omega$ $\mu m^2$ to M$\Omega$ $\mu m^2$, unless a MgO tunnel barrier is deposited before the CoFeB electrode, which lowers the resistance-area product below 10 k$\Omega$ $\mu m^2$.

## 6. Contacting 2D materials

After graphene, several two-dimensional materials have attracted growing interest for nano-electronic applications. In particular, the scientific community has rapidly understood the major limitation of graphene related to its lack of bandgap, which represents a fundamental property for the correct operation of a field effect transistor. Frenetic activity has therefore focused on the search for new two-dimensional materials [19,221]. Nowadays, more than 140 different 2D materials are known [222] and classified according to their structure. Silicene [223] and germanene [224] are representative of the X-enes family [225] in which also graphene, phosphorene [226] and stanene [227] are included, being the family of monolayer materials characterized by a single chemical element.

Transition metal dichalcogenides (TMDs) have chemical formula $MX_2$ (with M a transition metal element as Mo, W, Nb, etc. and X a chalcogen as S, Se or Te) and they are probably the most investigated 2D materials ($MoS_2$, $WS_2$, etc.) after graphene, due to their structural and electronic properties [228]. A monolayer is actually realized by three planes of atoms, with the M layer packed between two X layers. Due to the weak van der Waals forces holding together the layers, these materials can be easily peeled off in single layers as for graphene starting from graphite. The wide band gap (between 1 and 2 eV) makes TMDs suitable for nanoelectronic applications, such as field effect transistors and optoelectronic applications [225,229,230]. Different techniques allow to obtain $MX_2$ layers: micro-mechanical cleavage [231], CVD [232], or by precursor molecules spin coating [233]. The development of $MX_2$ based electronic devices is however strongly limited from the need to obtain low contact resistance at the $MX_2$/metal interfaces, as well. Indeed, the formation of states at the interface causes the pinning of Fermi level and the formation of significant Schottky barriers (in particular when p-type contacts are realized).

Heavy doping of semiconductor is a typical process to reduce the M/S contact resistance by lowering the Schottky barrier, but for 2D semiconductors it results very challenging [234-236]. One possibility is to cover the contacting metal by a layer to favor an increase of its work function. Indeed, metals typically realize n-type contacts with high contact resistance due to the formed Schottky barrier.

Despite the lack of dangling bonds in $MX_2$, metals with low workfunction cause the formation of interface states provoking pinning of Fermi level and an important (Schottky) barrier for the electrons [237-238].

Despite the large number, research on TMDs for use in electronic devices has mostly focused on molybdenum disulphide ($MoS_2$), although devices with $WS_2$, $WSe_2$, etc. have been reported [239-246]. The intense attention about $MoS_2$ arises from its natural availability and the high quality of 2D crystals that can be easily obtained. Single-layer $MoS_2$ is particularly interesting for electronic applications due to a direct band gap of 1.9 eV. $MoS_2$ is indeed an excellent candidate to obtain short-channel and gate length below 10 nm with $I_{ON}/I_{OFF}$ ratio above $10^8$ [247,248].

However, also for the exploitation of the electronic properties of these materials, including $MoS_2$, the formation of transparent contacts (low resistance) with three-dimensional metal electrodes is crucial to achieve significant device performance [249-251].

For instance, the n- or p-type behavior of $MoS_2$ FETs is controlled by charge injection from contacts, and ohmic contacts are crucial to characterize the intrinsic transport properties of the transistor channel. The high workfunction in Au, Ni, Pt favors n-type behavior [251], while both n-type and p-type behavior have been reported when using Pd electrodes on $MoS_2$ [252-254].

Several experiments already reported measurements of the contact resistance in FET structures realized with $MoS_2$ channel, and Ti/Au, Ni/Au or Au contacts [255-257]. The extracted values range between 0.2 and 2.0 k$\Omega\mu$m, still far from the best results obtained on graphene.

Several attempts to improve contact resistance at the interface metal-TMDs with different methods have been reported [238,239,255,258-261]. The most important step towards the improvement of contact resistance (and device performance) remains the reduction of the Schottky barrier at the metal-TMD interface.

An experimental study systematically tested different metals to contact $MoS_2$ from low work function metals, such as Sc and Ti (with workfunction 3.5 eV and 4.3 eV, respectively) to large work function metals (Ni and Pt, workfunction 5.0 eV and 5.9 eV, respectively) [259].

Baugher et al. [260] have demonstrated the possibility to obtain ohmic contacts using Ti/Au electrodes to the $MoS_2$ down to 4 K at high carrier densities by in situ vacuum annealing and electrostatic gating. They showed that vacuum annealing doped devices and reduces Schottky barriers and contact resistance.

A technique of phase engineered low-resistance contacts has been demonstrated by Kappera et al. [255] in $MoS_2$ FETs, where ultrathin semiconducting $MoS_2$ nanosheets are used as electrodes after local patterning of metallic 1T phase. With this procedure, contact resistances as low as 200 $\Omega\mu$m are obtained due to the atomically sharp interface. Importantly, device performance resulted reproducible and independent of the contact metal used.

An innovative solution to limit the extrinsic scattering factors and to achieve ultrahigh electron mobility has been proposed by Cui et al. [262] engineering a van der Waals heterostructure fully encapsulating the $MoS_2$ layer within hexagonal boron nitride and using graphene electrodes. More recently, low temperature ohmic contacts have been demonstrated in *h*-BN encapsulated TMDs by using a fabrication process based on selective etching of the top $h$-BN layer before depositing metal electrodes [263].

Another interesting approach to realize low-resistance contacts in $MoS_2$ based devices is to degenerately dope the contacts with potassium [264]. On the other hand, doping with elements that may react with the environment causes important instability as well as possible deterioration of the device with time.

Theoretical calculations have demonstrated that the effective masses of electrons and holes open fundamental challenges to obtain high switching speeds in $MoS_2$ based FETs [248].

Farmanbar et al. [265] have proposed a method to vary the Schottky barrier in p-type contacts by introducing a 2D buffer layer between the $MX_2$ semiconductor and the 3D metal electrode. The presence of two-dimensional van der Waals layer doesn't introduce interface states or any perturbation in absence of perfect lattice matching (see figure 18), while in case of direct metal/$MX_2$ contact, interface states are formed and Fermi level is pinned in the $MX_2$ band gap, originating a Schottky barrier. The use of a buffer layer suppresses the interaction at the interface, releasing the Fermi level; in such case $MX_2$ layer and buffer layer are bonded by van der Waals forces. As buffer layer, Farmanbar et al. [265] analyzed the possible application of graphene, monolayer of $h$-BN, or an oxide layer with a high electron affinity, such as $MoO_3$, or a metallic $NbS_2$ monolayer with a high work function, in order to realize zero Schottky barrier height for holes.

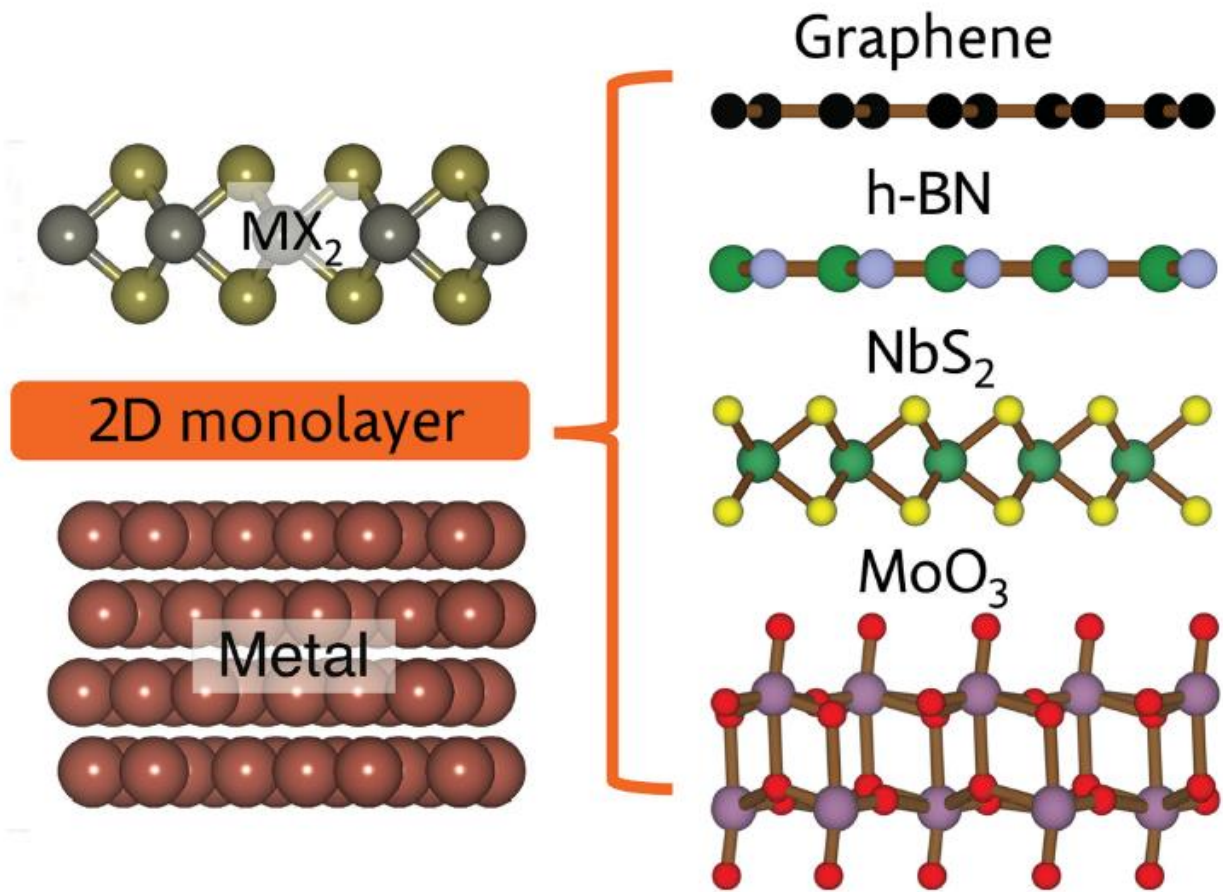

**Fig.18.** Buffer layer is introduced as interface between metal and $MX_2$ layer to favor low contact resistance. As possible buffer layers are reported graphene, *h*-BN, $NbS_2$, and $MoO_3$. The figure is reproduced from [265] with permission.

The calculations have shown that a bilayer of $MoO_3$ is effective as buffer layer to obtain a zero Schottky barrier height for contacts to any $MX_2$ semiconductor. However, the best solution to get zero barrier height is conclusively considered to be a metallic $NbS_2$ monolayer, having the additional advantage of high stability with respect to the strong oxidant $MoO_3$.

The formation of transparent electrical contacts remains presently one of the principal task to address in order to fully exploit the potentiality of 2D semiconductors.

## 7. Conclusions

The problem of low contact resistance to graphene is of relevant technological interest, being a crucial factor to exploit the extraordinary graphene electronic properties in high performance devices. Due to the hectic activity in the community involved more generally in the comprehension and application of the 2D materials, graphene has a central role, being the first discovered material with incomparable properties and envisaging future nanoelectronics based on flexible electrodes as well as photovoltaic and sensing devices.

The understanding of graphene-metal interface is therefore of fundamental importance, the contact regions between the 2D graphene and the 3D metal electrodes exhibiting peculiar properties with respect the usual metal-semiconductor contacts. Indeed, metal contact causes significant doping of graphene and Fermi level shift.

Since charge injection at the interface is relevant to the contact resistance, several procedures have been proposed and/or experimentally tested in order to enhance carrier transmission and lower the contact resistance.

Here, we reviewed the wide scientific activity dealing with the improvement of contact resistance through innovative fabrication process of metal contacts as well as surface treatments and formation of one-dimensional edge contacts of graphene to three-dimensional metal electrodes. The future of graphene and other two-dimensional materials in the field of nanoelectronics with successful applications will be severely related to the complete understanding and control of the interfacial phenomena as the contact resistivity.